\documentclass[12pt]{article}
\usepackage{amsmath, amsfonts}
\newcommand{\bee}{\begin{eqnarray*}}
\newcommand{\ene}{\end{eqnarray*}}
\newcommand{\beeq}{\begin{equation}}
\newcommand{\eneq}{\end{equation}}
\newtheorem{lem}{Lemma}[section]
\newcommand{\bel}{\begin{lem}}
\newcommand{\enl}{\end{lem}}
\newtheorem{defi}{Definition}[section]
\newcommand{\bef}{\begin{defi}}
\newcommand{\enf}{\end{defi}}
\newtheorem{exap}{Example}[section]
\newcommand{\beex}{\begin{exap}}
\newcommand{\enex}{\end{exap}}
\newtheorem{theo}{Theorem}[section]
\newcommand{\beth}{\begin{theo}}
\newcommand{\enth}{\end{theo}}
\newtheorem{prop}{Proposition}[section]
\newcommand{\bep}{\begin{prop}}
\newcommand{\enp}{\end{prop}}
\newtheorem{cor}{Corollary}[section]
\newcommand{\bec}{\begin{cor}}
\newcommand{\enc}{\end{cor}}
\newtheorem{rem}{Remark}[section]
\newcommand{\ber}{\begin{rem}}
\newcommand{\enr}{\end{rem}}

\usepackage{amsmath}
\usepackage{graphicx}
\usepackage{latexsym}
\begin{document}
\title{ 
A NEW METHOD TO OBTAIN\\
RISK NEUTRAL PROBABILITY,\\
WITHOUT STOCHASTIC CALCULUS\\  AND PRICE MODELING,\\
CONFIRMS  THE UNIVERSAL VALIDITY\\
OF  BLACK-SCHOLES-MERTON FORMULA\\
AND
VOLATILITY'S ROLE\\
}
\author{Yannis  G. Yatracos\\
Cyprus U. of Technology
}
\maketitle
\date{}

{\it e-mail:} yannis.yatracos@cut.ac.cy

\pagebreak
 
\begin{center}
{\bf Summary}
\end{center}

A new method is proposed to obtain the risk neutral probability of  share prices
without stochastic calculus and price modeling, via an 
embedding of the
 price return modeling problem in Le Cam's
statistical experiments framework.
Strategies-probabilities $P_{t_0,n}$ and $P_{T,n}$ are thus determined and used, respectively, 
for the trader selling the share's European call option at time
$t_0$  and for the buyer who may exercise it in the future, at $T; \  n$ increases with the number 
of share's transactions in $[t_0,T].$
When the transaction times are dense in $[t_0,T]$
it is shown, with mild conditions,
that
under each of these probabilities $\log \frac{S_T}{S_{t_0}}$ has
infinitely divisible distribution
and in particular normal distribution for ``calm''
share; $S_t$ is the share's price at time $t.$
The price of the share's 
call is the limit of the expected values of the call's  payoff
under the translated 
$P_{t_0,n}.$
A formula for the price is obtained. It coincides for the special case of  ``calm'' share prices
with the Black-Scholes-Merton 
{\em formula} with variance not necessarily  proportional to $(T-t_0),$  thus confirming
formula's 
universal validity without model assumptions.
Additional results clarify volatility's role in the transaction and the behaviors of the trader 
and the buyer.
Traders may use the pricing formulae after estimation of the unknown parameters.

\vspace{1in}


{\em Running Head:}
Obtaining risk neutral probability and applications

\pagebreak



\section{\bf Introduction}

\quad In this work,  Le Cam's theory of statistical experiments is used to obtain in two steps the risk neutral probability
${\cal P}^*$  of a share's price 
  and to price the share's European call option without 
stochastic calculus and price modeling.
Among other results, the universal validity of the Black-Scholes-Merton ({\em B-S-M}) formula (Black and Scholes (1973), and Merton (1973)) 
is confirmed for  ``calm'' stock prices, a pricing formula is obtained for non-calm stock 
 and new insight is provided for volatility's role and for the behaviors of the trader
 selling the call and of the  buyer.

 The buyer,  at time $t_0,$  of  a share's European call  has  at time $T (>t_0)$ the option to buy the share at
predetermined 
 price $X.$  The
``fair''  {\em B-S-M}- price $C$  is obtained
when  the 
share's prices $\{S_t, t>0\},$ defined  on the probability space $(\Omega, {\cal F},{\cal  P}),$  follow a Geometric Brownian motion.
However,  $C$  is
 used extensively by traders  for various  price models 
 because of its simplicity, adjusting  $C$  for the demand that reflects market's expectations.


{\em  B-S-M}  price $C$ is obtained
by ``replicating the call,''  i.e. by creating a portfolio 
that matches the call's
payoff at $T.$
 This procedure guarantees $C$ does not allow arbitrage, 
i.e. that  the call option's buyer  cannot  make profit with probability 1.
Alternatively,
$C$ is obtained by discounting at $t_0$ the expected value of the call's payoff,
$(S_T-X)I(S_T>X),$ 
 under  the risk neutral  probability  ${\cal P}^*$ that is equivalent to  physical probability ${\cal P},$
for which the discounted shares'prices $\{\tilde S_t=e^{-rt}S_t, t>0\}$ form a martingale; $I$ denotes
the indicator function,  $r=\ln(1+i), \ i$ is the fixed interest
(see, for example, Musiela and Rutkowski, 1997).
Stochastic calculus is used with both approaches to obtain $C$ via $P^*$ {\em but} 
 in practice
${\cal P}$ and $S_t$'s law under ${\cal P}$ are  not known  {\em and} ${\cal P}^*$ is not always  easily  obtained.

 Cox, Ross and Rubinstein (1979) 
obtained for the binomial  price model 
 {\em B-S-M} price  as limiting price when the model's
parameters are properly chosen.
This result,  the form of  $C$ which   indicates the existence of contiguous sequences of probabilities and
 its  extended use
all suggest  that the
``fair'' option's price 
may  be obtained for various  price models as 
limit of  expected values
under   
a sequence of probabilities.
The conjecture is  proved herein without stock price modeling assumptions.

In section 2 a new 2-step method is proposed to determine ${\cal P}^*$ without model assumptions for the share price  via an
embedding in Le Cam's statistical experiments
framework. 
The tools, presented in section 3,  include mean-adjusted  prices $S_t/ES_t, \ t\ge t_0,$ which are densities in   
$(\Omega, {\cal F}, {\cal P})$ but have also mean 1 independent of $ES_t.$
These are used to define two strategies-probabilities $P_{t_0,n}$ and $P_{T,n},$ respectively, for
the trader who sells the call at $t_0$ and for the buyer, 
 through the $k_n$ traded stock prices at times $t_0, \ t_1, \ldots, t_{k_{n-1}}$  in  the  interval $[t_0,T), 
\  T=t_{k_n}.$ 
These probabilities
are used
to derive the distribution of $\log \frac{S_T}{S_{t_0}}$
when $k_n$ increases to infinity (with $n$)
and price the call using  $P_{t_0,n}$   properly translated.

In modeling section 4
it is shown,  with mild conditions, that
when the non-random transaction times are dense in $[t_0,T]$
the  distributions of $\log \frac{S_T}{S_{t_0}}$ under 
$P_{t_0,n}$ and $P_{T,n}$ 
are infinitely divisible (Theorem  \ref{t:infdiv}). 
The prices-densities $\{S_t/ES_t, \ t=t_0^n,\ t_1^n, \ldots , t_{k_n}^n\}$
form a martingale under
$P_{t_0,n}$ (Theorem  \ref{t:MG});  $t_0^n=t_0, \ t_{k_n}^n=T$ for each $n.$
For calm stock prices, with prices-densities not changing much often,
both distributions of $\log \frac{S_T}{S_{t_0}}$ are normal (Theorem  \ref{t:asydicalm})
and the role of volatility in the
transaction
 is confirmed 
since the statistical experiment
${\cal E}_n=\{P_{t_0,n},  \ P_{T,n}\}$
converges to the Gaussian experiment
${\cal G}=\{\tilde P=N(0,1), \  \tilde Q=N(\sigma_{[t_0,T]},
1)\}$ (Corollary  \ref{c:convexp});
$\sigma_{[t_0,T]}$ is the standard deviation  associated with
the
distribution of $\log \frac{S_T}{S_{t_0}}.$
Under $\tilde P,$ $\log \frac{d\tilde Q}{d\tilde P}$  follows a normal distribution
with mean $-.5 \sigma^2_{[t_0,T]}$ and variance $\sigma^2_{[t_0,T]}.$
Conditions are also provided under which $\sigma^2_{[t_0,T]}=
\sigma^2(T-t_0), \ \sigma>0$ (Proposition  \ref{p:T-t}).



In section 5,
prices of the  call  are obtained 
for the trader and the buyer.
 These prices are limits
 of discounted at $t_0$   
expected values of
the call's payof
 under $P_{t_0,n}$ and  $P_{T,n},$ respectively,
 as  $n$  increases to infinity.
Without loss of generality and to reduce the difficulty in the presentation $S_T$ is discounted with $ES_{t_0}/ES_T$ and 
$X$
with $e^{-r(T-t_0)}$  but the discounting factors coincide when the limit 
distribution becomes risk neutral using translation; recall that under $P^*,$ $e^{-r(T-t_0)}=s_{t_0}/E_{P^*}S_T$
(Corollaries  \ref{c:pri1},  \ref{c:pri2}).
Using a mapping to an equivalent risk-neutral structure, obtained by translation
or change of probability $P_{t_0,n}$
 conditional on 
the value of the compound Poisson component of 
$\log \frac{S_T}{S_{t_0}},$ 
the 
additive term $\log \frac{ES_T}{ES_{t_0}}$ that appeared due to $S_T$'s  discounting is replaced by $r(T-t_0).$
For calm stock this mapping
 leads to the
{\it B-S-M} formula with  $\sigma^2_{[t_0,T]}$ not necessarily proportional to 
$(T-t_0)$  thus proving its universal validity (Corollaries  \ref{c:pri1calm}, 
 \ref{c:pri2calm})
and justifying its  frequent use for various  price models
that has puzzled  Musiela and Rutkowski (1997, p. 111. l. -7 to l. -1).
For non-calm stock, the obtained
price has the same constants as the {\it B-S-M} price
but the probabilities (in {\it B-S-M}) are now integrands
with respect to the probability of the Poissonian component
in the distribution of $\log \frac{S_T}{S_{t_0}}.$
It  is reminiscent of  
the price in Merton (1976, p. 127)
 obtained 
when the stock's price 
consists of
 {\it (i)} the ``normal'' vibration,
modeled by a standard geometric
Brownian motion, and {\it (ii)} the ``abnormal'' vibration,
due for example to important new information about the stock, that is
modeled by
a jump process ``Poisson driven''.
The trader can use the obtained prices by estimating the unknown parameters.


Additional results  clarify and confirm
quantitatively for calm stock that:\\
{\it a)} the price the buyer is expected to pay for the option 
includes
indeed a volatility premium (Corollaries  
 \ref{c:pri1calm} {\it (ii)}, 
 \ref{c:pri2calm}),
and\\ 
{\it b)} the probability $S_T$ is greater than 
the strike price $X$ is larger
for the buyer than for the trader (Theorems  \ref{t:pri1termscalm}, 
{\it (i), (ii)},  \ref{t:pri2termscalm}, {\it (i), (ii)}).\\
Similar results hold for non-calm stock.

The results in this work hold
 also for random interest rate; see Remarks  
\ref{r:stochint1},  \ref{r:stochint2}.


 Fama's weak Efficient Market Hypothesis  implies 
either independence or slight dependence of the share price returns
(Fama, 1965,  p.90, 1970, pp. 386, 414). Modeling ``slight" dependence with  weak dependence  is acceptable in  Finance (Duffie, 2010, personal communication). Thus, limiting laws  obtained 
under independence of the price returns (used in $P_{t_0,n}$)  remain valid under weak dependence and the same holds 
for the obtained prices.

The results in this work and results in Yatracos (2013) relating {\em B-S-M} price with Bayes risk and applications have been presented since 2008  in various seminars and in particular in  the Third International Conference on 
Computational and Financial Econometrics (2009, http://www.dcs.bbk.ac.uk/cfe09/).
Janssen and Tietje (2013) used also Le Cam's theory of statistical experiments to discuss ``the connection between mathematical
finance and statistical modelling''(see the Summary) for $d$-dimensional price processes.
 Known results from statistical experiments are used in order to revisit financial models (p. 111, lines 22, 23).
Some of the differences in their work are: {\em a)} The  price process  is not standardized and a risk neutral
probability is assumed to exist. {\em b)} Convergence of the likelihood ratios to a normal experiment is obtained under the assumption of contiguity. 
{\em c)} There are no results explaining the behaviors of the trader and of the buyer.
{\em d)} There is no proof of the universal validity of 
{\em B-S-M} formula  without modeling assumptions on the shares' prices.

The theory of statistical experiments 
used 
is in Le Cam (1986, Chapters 10 and 16),
Le Cam and Yang (1990, Chapters 1-4, 2000, Chapters 1-5)
 and in Roussas (1972, Chapter 1).
A concise introduction in this material can be found 
in Pollard (2001). Theory of option pricing can be found, among others, in
Lamberton and Lapeyre (1996) and Musiela and Rutkowski (1997).
Proofs are in the Appendix.

\section{A method to obtain risk neutral  probability}

\quad   The price of a European  option  is independent from the 
stock prices' drift  and depends only on the risk free interest rate. 
This is usually attained by finding a risk neutral probability ${\cal P}^*$ equivalent to the physical probability ${\cal P}$ for which 
\begin{equation}
\label{eq:weneed}
E_{{\cal P} ^*}(\frac{S_T}{S_t}|{\cal F}_t)=e^{r(T-t)}; 
\end{equation}
$\{ {\cal F}_t\}$ is the natural filtration. For example, 
 when  stock prices are  log-normal  
Girsanov's theorem is used to obtain  ${\cal P}^*$  
and the drift  disappears obtaining (\ref{eq:weneed}).

For several stock price models  it is not easy to determine  ${\cal P}^*.$ 
Moreover, the stock-price model is {\em usually  unknown}. How one can  proceed  in this situation?
We can  try to  obtain (\ref{eq:weneed}) with a different approach in two steps. 
Observe that  the adjusted prices $\{S_t/ES_t\}$  are independent of the drift since they all have mean 1; $ES_t$ is taken under ${\cal P}.$  

{\bf Step 1:} Decompose $\log (S_T/S_t)$ in two components,
\begin{equation}
\label{eq:motiv}
\log \frac{S_T}{S_t}= \log \frac{S_T/ES_T}{S_t/ES_t}+ \log \frac{ES_T}{ES_t}.
\end{equation}
Determine a probability $Q$ under which
$S_t/ES_t$
is a martingale.
There is no involvement of the interest (i.e. of  $r$)  in this step.

{\bf Step 2:} Use  $Q$ defined in step 1 and 
 translate $\log \frac{S_T}{S_t}$  by $r(T-t)- \log \frac{ES_T}{ES_t}$ thus  obtaining  ${\cal P}^*$ 
under which   $\frac{S_T}{S_t}$  satisfies  (\ref{eq:weneed}).
Probabilists prefer to obtain ${\cal P}^*$ with  a change of probability $Q$ via $\frac{d{\cal P}^*}{dQ}.$  

 This new approach works for log-normal prices and allows to obtain the B-S-M price without stochastic calculus. 

\beex
\label{ex:lognormBSM}
 Let $S_t$ be a geometric Brownian motion,
\begin{equation}
S_t=s_0exp\{(\mu-\frac{\sigma^2}{2})t+\sigma B_t\}
\label{eq:gbm}
\end{equation}
with $B_t$ standard Brownian motion, $t>0$ and $s_0$ the price at
$t=0.$
For $t<T,$
\begin{equation}
\log \frac{S_{T}}{S_{t}}=(\mu-\frac{\sigma^2}{2})(T-t)+
\sigma (B_{T}-B_{t}),
\label{eq:crudeBS}
\end{equation}
and since $ES_t=s_0exp\{\mu t\},$
$S_t/ES_t$  is a martingale under ${\cal P}$ (i.e. $Q={\cal P}$)  and you can obtain (1) 
via (\ref{eq:motiv}) either simply translating $\log \frac{S_T}{S_t}$  by $r(T-t)- \log \frac{ES_T}{ES_t}$   or with a change of probability  via Lemma \ref{l:probchange},
$$\frac{d{\cal P}^*}{d{\cal P}}(w)=e^{Aw+C},$$
 with  
$$A=\frac{r-\mu}{\sigma^2}, \ C=-MA-\frac{\Sigma^2A^2}{2},\  M=(\mu-\frac{\sigma ^2}{2})(T-t), \  \Sigma^2=\sigma^2 (T-t). $$
It follows that  under ${\cal P}^*$ the law of $\log \frac{S_T}{S_t}$ is normal with mean
$M^*=M+A\Sigma^2=(r-\frac{\sigma ^2}{2})(T-t)$ and variance $\Sigma^{*2}=\Sigma^2=\sigma^2(T-t)$ and
(\ref{eq:weneed}) holds.
\enex

In Example \ref{ex:lognormBSM} ${\cal P}^*$ is easily obtained  because the distribution of $\log \frac{S_T/ES_T}{S_t/ES_t}$ is normal.  Can one similarly obtain $Q$ and $P^*$  in other situations? 
Note that without share price model assumption we have only the observed share prices. It is similar to a non-parametric statistical problem where the observations have all the information.  {\em The share  prices should determine the 
``right" $Q$ or
a sequence $Q_n$} that will give us the distribution of   $\log \frac{S_T/ES_T}{S_t/ES_t}$ with an asymptotic argument via a 
Central Limit Theorem. The word ``asymptotic'' is used in the sense that there are countably  infinite many  transactions in  each sub-interval of  $(t,T).$

 When there are  ``many'' transactions in  $[t_0,T],$  an embedding in Le Cam's statistical experiments 
allows to determine the  sequence of probabilities $Q_n=P_{t,n}$  under which  
$\frac{S_T}{S_t}$ 
satisfies (\ref{eq:weneed}) via the translation in  Step 2 for the law of $\log \frac{S_T/ES_T}{S_t/ES_t}$ under $Q_n.$

\section{The embedding, the tools and  prices-densities}

\quad Let $(\Omega, {\cal F}, {\cal P})$ be the underlying probability space
of the stock prices $\{0<S_t, \ 0\le t \le T\},$
and let $ES_t$ be the expectation with respect to ${\cal P}.$
Consider 
the process of
prices-densities 
\begin{equation}
\{p_t=\frac{S_t}{ES_t}, \ t \in [0,T]\}.
\label{eq:pricedensities}
\end{equation}

Define on  $(\Omega, {\cal F}, {\cal P})$ the 
(forward) probability $P_t:$ 
for $A \in {\cal F}$
\begin{equation}
P_t(A)=\int_A \frac{S_{t}(\omega)}{ES_{t}} {\cal P}(d\omega);
\label{eq:stockprob}
\end{equation}
the derivative of $P_t$ with respect to $P,$
\[
\frac{dP_t}{d{\cal P}}=\frac{S_t}{ES_t}=p_t, \ t \in [0,T].
\]
Since $S_t$ is positive, $P_t$ and
${\cal P}$ are mutually absolutely continuous. Thus,
$P_t$ and $P_s$ are mutually absolutely continuous
for each $t,  \ s$ in $\Theta=[0,T],$ and
\[
\frac{dP_t}{dP_s}=\frac{p_t}{p_s} \ a.s. \ {\cal P}.
\]

A binary statistical experiment ${\cal E}=\{P, \ Q\}$ with
$P,\ Q$  probabilities on $(\tilde \Omega, \tilde {\cal F})$ (Blackwell, 1951).
Le Cam (see, for example, 1986) defined ${\cal E}$ as Gaussian experiment 
when $P$ and $Q$ are equivalent and the distribution of
$\log \frac{dQ}{dP}$ under either $P$ or $Q$ is normal, 
 introduced 
a distance $\Delta$ 
between experiments 
and
proved that $\Delta$-convergence of experiments
${\cal E}_n=\{P_n, Q_n\}, \ n\ge 1,$
to ${\cal E}$
is equivalent to
weak convergence of likelihood ratios
$\frac{dQ_n}{dP_n}$ under $P_n$ (resp. $Q_n$)  to the distribution
of $\frac{dQ}{dP}$ under $P$ (resp. $Q$).
In analogy with the frequent
weak convergence of sums of
random variables to a Gaussian distribution, there is
frequent $\Delta$-convergence
of experiments to a Gaussian experiment.

Embed
the traded stock prices in the
statistical 
experiments framework
via $\log \frac{dQ_n}{dP_n}$
by re-expressing $ \log \frac{S_T}{S_{t_0}}$ as in (\ref{eq:motiv}) using also the intermediate prices,
\begin{equation}
\log \frac{S_T}{S_{t_0}}=\log \frac{\frac{S_T}{ES_T} \ldots \frac{S_{t_1^n}}
{ES_{t_1^n}}}
{\frac{S_{t_0}}{ES_{t_0}} \ldots \frac{S_{t^n_{k_n-1}}}{ES_{t^n_{k_n-1}}}}
+\log \frac{ES_T}{ES_{t_0}}.
\label{eq:embedding}
\end{equation}

The products of normalized prices-densities 
$\frac{S_{t_0}}{ES_{t_0}} \ldots \frac{S_{t^n_{k_n-1}}}{ES_{t^n_{k_n-1}}}$ and
$ \frac{S_T}{ES_T} \ldots \frac{S_{t_1^n}}
{ES_{t_1^n}}$
determine, respectively, $P_n$ and  $Q_n$ in $(\Omega^{k_n}, {\cal F}^{k_n}).$
Le Cam's theory
provides
the asymptotic distribution of $\log \frac{dQ_n}
{dP_n}$ under $P_n.$
The trader 
 uses $P_n$ denoted in the sequel by $P_{t_0,n}.$
Naturally, the buyer who acts one
transaction period later than the trader uses $Q_n,$ denoted $P_{T,n}.$
To calculate the option's prices
the laws
of $\log \frac{S_T}{S_{t_0}}$ 
under $P_{t_0, n}$ and $P_{T, n}$ are used
when 
$k_n \rightarrow \infty.$  To obtain the ``fair'' price  the limit  law under $P_{t_0, n}$  will be translated such that 
(\ref{eq:weneed}) holds.


Consider on $(\Omega^{k_n}, {\cal F}^{k_n})$ the statistical experiment
\begin{equation}
{\cal E}_{k_n}=\{P_{t_0,n}=\Pi_{j=0}^{k_n-1}P_{t_j^n}, \mbox{   }
P_{T,n}=\Pi_{j=1}^{k_n}P_{t_j^n} \},
\label{eq:statexp1}
\end{equation}
with $t_0^n=t_0$ and $t_{k_n}^n=T$ for each $n.$  
$P_{t_0,n}$ is determined  
via (\ref{eq:embedding}) 
 by 
\begin{equation}
P_{t_0,n}(B_0x \ldots x B_{k_n-1})=P_{t_0}(B_0)P_{t_1^n}(B_1) \ldots
P_{t^n_{k_n-1}}(B_{k_n-1}),
\label{eq:prodmeas}
\end{equation}
for $B_{j} \in {\cal F}, \ j=0,\ldots, k_n-1,$ and its extension to
the product $\sigma$-field ${\cal F}^{k_n}.$

$P_{T,n}$ is determined similarly by prices-densities 
at $t_1^n, \ldots,
t_{k_n}^n.$ 
 $P_{t_0,n}$ and $P_{T,n}$
are  mutually absolutely continuous since
$S_t>0$ when $t>0.$

Let
\begin{equation}
Y_{n,j}=\sqrt { \frac {p_{t_j^n}}{p_{t^n_{j-1}}} }-1, \hspace{4ex}
U_{n,j}=\sqrt { \frac {p_{t_{j-1}^n}}{p_{t^n_j}} }-1, \ j=1,\ldots, k_n,
\label{eq:yu}
\end{equation}
\begin{equation}
a[t_0,T]=\frac{ES_T}{ES_{t_0}},
\label{eq:a}
\end{equation}
\begin{equation}
\Lambda_{k_n}=\log \frac{p_T}{p_{t_0}}=\log \frac{\Pi_{j=1}^{k_{n}}
p_{t_j^{n}} }
{\Pi_{j=1}^{k_{n}}
p_{t_{j-1}^n} }=2 \sum_{j=1}^{k_n} \log(1+Y_{n,j})
=-2 \sum_{j=1}^{k_n} \log(1+U_{n,j}).
\label{eq:distribution}
\end{equation}
Rewrite $\log (S_T/S_{t_0})$ using (\ref{eq:distribution})
via (\ref{eq:embedding}) and (\ref{eq:a}):
\begin{equation}
\log (S_T/S_{t_0})=\Lambda_{k_n} + \log a[t_0,T].
\label{eq:logratio}
\end{equation}
Experiment (\ref{eq:statexp1}) is not specified since the prices-densities
and therefore both
$P_{t_0, n}$ and $P_{T, n}$ are all unknown but, under mild conditions,
it is shown using Le Cam's theory and (\ref{eq:logratio}) that
when $k_n$ increases to infinity
$\log \frac{S_T}{S_{t_0}}$ has infinitely divisible distributions
under $P_{t_0,n}$ and $P_{T, n}.$ 



\bep \label{p:Yaugm}
Define on $(\Omega^{k_n},
{\cal F}^{k_n})$ random variables
$$\tilde Y_{n,j}(\omega^{(k_n)})=Y_{n,j}(\omega_j), \ j=1,\ldots, k_n,$$
$\omega^{(k_n)}=(\omega_1,\ldots, \omega_{k_n}).$
The variables 
$\tilde Y_{n,j}, \ j=1,\ldots,k_n,$ are independent
under $P_{t_0,n}$ and $P_{T,n},$ 
i.e. $Y_{n,j}(\omega_j)$ is independent
of $Y_{n,k}(\omega_k), \ j \neq k.$
The same holds for  $U_{n,j}, \ j=1,\ldots,k_n,$ and 
$\frac{p_{t_j^n}}{p_{t_{j-1}^n}}, \ j=1,\ldots,k_n.$
\enp

\ber \label{r:Yaugm} Proposition \ref{p:Yaugm} is used to derive the 
asymptotic distribution of $\Lambda_{k_n}$ (see \ref{eq:distribution}) under
$P_{t_0,n}$ and under $P_{T,n}$ via $\{\tilde Y_{n,j}, \ j=1,\ldots, k_n \}$
when $n \rightarrow \infty.$
\enr

Prices-densities and their probabilities 
with respect to ${\cal P}$ have been already used
in the Finance literature to express wealth in a new
num\'{e}raire;  see, for example, Detemple and Rindisbacher (2008)  
who attribute the notion of price-density and the obtained
``forward probability measure'' (like $P_t$ in (\ref{eq:stockprob}))
to Geman (1989) and Jamshidian (1989).

The use of prices-densities and of $P_{t_0,n}$ and $P_{T,n}$
is now motivated from different angles.


{\it (i)} Assume that stock prices have all the information.
Just prior to $t_0^n$
the (unobserved) information and in particular the volatility
in $(S_{t_0^n},\ldots,
S_{t_{k_n}^n})$
is better measured by their coefficient of variation that is the variance
of the corresponding price-densities $(p_{t_0^n}, p_{t_1^n},
\ldots, p_{t^n_{k_n}}).$ The trader has the information
from $t_0^n=t_0$ until $t_{k_n-1}^n,$ and the buyer has the information 
from $t_1^n$ until $t_{k_n}=T$ expressed respectively by $(p_{t_0^n},\ldots, 
p_{t^n_{k_n-1}})$ and $(p_{t_1^n},\ldots, p_{t^n_{k_n}}).$ 
If both have similar information
the transaction will probably occur. 
These information vectors are  embedded in $(\Omega^{k_n}, {\cal F}^{k_n})$ 
obtaining 
probabilities
$P_{t_0,n}$ and $P_{T,n}.$  
They both miss one component from the whole information 
$(p_{t_0^n}, p_{t_1^n}, \ldots, p_{t^n_{k_n}})$ but this additional
information becomes negligible when $k_n$ is large,
at least in some situations (Mammen, 1986).

{\it (ii)} To compare different stocks (or assets) 
Modigliani and Miller (1958) used the ratio of all long term debt
outstanding to the book value of all common stock, and called it
``leverage'' (or ``financial risk'')  measure.
To compare values of the same stock Sprenkle (1961) used as leverage
ratios of successive prices $\frac{S_{t_j^n}}{S_{t_{j-1}^n}}$
without expectation normalization
and obtained the price of a call option for a warrant that is similar
to the {\it B-S-M} price with
arbitrary multiplicative parameters $k=\frac{ES_T}{s_{t_0}}$
and $k^*$  preceding,respectively,
$\Phi(d_1)$ and $\Phi(d_2)$ (see form of  (\ref{eq:bsmspr}));  $k^*$ is a discount factor that
depends on the risk of the stock 
(Black and Scholes, 1973, p. 639). In Sprenkle's ratios, variances 
are not adjusted for the price 
level
and the obtained price is not the {\it B-S-M} price.
Using instead  $Y_{n,j}$ and $U_{n,j}$ in (\ref{eq:yu})
to measure financial risk,
the {\it B-S-M}
price is obtained for calm stock after translation of
$\log \frac{S_T}{S_{t_0}}.$



\section{Modeling the distribution of $\log \frac {S_T}{S_{t_0}}$}

\subsection{The Modeling Assumptions}

\quad Let $S_t$ be
the stock value at time $t, \ t\in [0,T],$
defined
on $(\Omega, {\cal F}, P).$ 
Assume 

${({\cal A}1)}$  $S_t>0$ and $ES_t<\infty$ for
every $t \in [t_0,T],$

${({\cal A}2)}$  a countable number of transaction times
in any open interval
of $[t_0,T],$

${({\cal A}3)}$ for 
the prices-densities $p_{t^n_0}, \ p_{t_1^n}, \ldots, p_{t^n_{k_n}}$
with mesh size $\delta_n=\sup \{t_j^n-t^n_{j-1}; j=1,\ldots, k_n\}, \
k_n=k_n(\delta_n),$ 
\[
(i) \lim_{\delta_n\rightarrow 0}
\sup \{ E_{P_{t^n_{j-1}}}(\sqrt{\frac{p_{t_j^n}}{p_{t_{j-1}^n}}}-1)^2,
\ j=1,\ldots, k_n\}=0,
\]
\[
(ii) \sup_n\sum_{j=1}^{k_n}
E_{P_{t^n_{j-1}}}(\sqrt{ \frac{p_{t_j^n}}{p_{t_{j-1}^n}}}-1)^2  \le b<\infty.
\]

Assumption ${{\cal A}1}$ allows in our framework
the passage from stock prices to prices-densities.
Assumptions ${{\cal A}1}$ and ${{\cal A}2}$ provide
via (\ref{eq:stockprob})
strategies $P_{t_0,n}$  and $P_{T,n}.$
Assumption ${\cal A}3 (i)$ indicates that 
the contribution of the ratio 
$\frac{p_{t_j^n}}{p_{t_{j-1}^n}}$
does not affect the distribution of $\frac{p_{T}}{p_{t_0}}, \ j=1,\ldots,k_n.$ 
Assumption ${\cal A}3 (ii)$  implies 
that the sum of the variances of $Y_{n,j}, 
\ j=1,\ldots,k_n,$ (see (\ref{eq:yu})) is uniformly bounded.

\ber
For the expectations in ${\cal A}3$ it holds
$$E_{P_{t^n_{j-1}}}(\sqrt{ \frac{p_{t_j^n}}{p_{t_{j-1}^n}}}-1)^2
=\int (\sqrt{ p_{t_j^n}}-\sqrt{  p_{t_{j-1}^n} })^2 d{\cal P}
\propto H^2(P_{t_j^n},
P_{t_{j-1}^n});$$
$H(P_{t_j^n},P_{t_{j-1}^n})$ is Hellinger's distance of 
$P_{t_j^n}, P_{t_{j-1}^n}$ defined in (\ref{eq:defhdxi}).
\enr


\subsection{\bf Infinitely divisible distribution of $\log \frac{S_T}{S_{t_0}}$
}

\quad Assumption ${\cal A}3$ implies 
the sequences of distributions of $\sum_{j=1}^{k_n}Y_{n,j}$ 
and of $\sum_{j=1}^{k_n}U_{n,j}$  are each relatively 
compact both under $P_{t_0,n}$ and $P_{T,n}.$ 
Thus, we can choose a subsequence $\{k_{n'}\},$  
for which both $\sum_{j=1}^{k_{n'}}Y_{n',j}$ and 
$\sum_{j=1}^{k_{n'}}U_{n',j}$ 
converge weakly, respectively, 
under $P_{t_0,n'}$ and $P_{T,n'}.$
Without loss of generality we will use $\{n\}$ and $\{k_n\}$
instead of $\{n'\}$ and $\{k_{n'}\}.$
The next result
determines the distribution of $\log \frac{S_T/ES_T}{S_{t_0}/ES_{t_0}}$ from
its moment generating function of an infinitely divisible distribution.
Recall that any infinite divisible distribution is that of the sum of two independent components, one normal and one Poissonian, and these components are unique up to a shift (Meerschaert and Scheffler,  p.41).  

\beth \label{t:LLCmain}
(Le Cam, 1986, Proposition 2, p. 462) Assume that ${\cal A}3$ holds.
Then, $\Lambda_{k_n}=\log \Pi_{j=1}^{k_n}\frac{p_{t_{j}^n}}{p_{t_{j-1}^n}}$ 
converges 
under 
$P_{t_0,n}$ in distribution to $\Lambda_{t_0}$ such that for
every $s \in (0,1)$ it holds
\begin{equation}
\log Ee^{s\Lambda_{t_0}}
=(2\mu-\sigma^2)s+2\sigma^2s^2+\int_{[-1,0)\cup (0,\infty)}
[(1+y)^{2s}-1-2sy]L_{t_0}(dy),
\label{eq:mgflambda}
\end{equation}
where $\mu=\lim_{n \rightarrow \infty}\sum_{j=1}^{k_n}E_{P_{t_{j-1}^n}}Y_{n,j},$
\\ $\sigma^2=
\lim_{\tau \rightarrow 0} \lim_{n \rightarrow \infty} \sum_{j=1}^{k_n}
E_{P_{t_{j-1}^n}}Y_{n,j}^2I(|Y_{n,j}|\le \tau),$\\
the L\'{e}vy measure \\ $L_{t_0}(x)=
\lim_{n \rightarrow \infty}L_{k_n}(x)
=\lim_{n \rightarrow \infty}\sum_{j=1}^{k_n}
E_{P_{t_{j-1}^n}}Y_{n,j}^2I(Y_{n,j}\le x)$\\
and $L^*_{t_0}$ is the probability determined by the 
Poissonian
component of $\Lambda_{t_0}.$
\enth

From (\ref{eq:mgflambda}) the normal
component of $\Lambda_{t_0}$
has mean $\mu_{[t_0,T]}$ and variance  $\sigma^2_{[t_0,T]},$ with
$$\mu_{[t_0,T]}=2\mu-\sigma^2, \hspace{4ex} \sigma^2_{[t_0,T]}=4\sigma^2.$$
The mean $\mu_{[t_0,T]}$ is determined by variances since by ${\cal A}3$
$$\mu_{[t_0,T]}=-\lim_{n\rightarrow \infty}\sum_{j=1}^n
Var(Y_{n,j})-\sigma^2.$$

\bec \label{c:lambda}
When ${\cal A}3$ holds, the asymptotic distribution of $-\Lambda_{k_n}$
under $P_{T,n}$ has the same $\mu$ and $\sigma^2$ as $\Lambda_{t_0}$
but different L\'{e}vy
measure. Under $P_{T,n},$ $\Lambda_{k_n}$  converges weakly to $\Lambda_T,$
with shift component $-\mu_{[t_0,T]}=-(2\mu-\sigma^2),$ 
$\sigma^2_{[t_0,T]}=4 \sigma^2$ and L\'{e}vy
measure $L_T.$ 
The 
Poissonian 
component of $\Lambda_T$ has probability $L^*_T.$
\enc
The distribution of $\log \frac{S_T/ES_T}{S_{t_0}/ES_{t_0}}$ follows from  next 
theorem.  The risk neutral probability
 follows
from (\ref{eq:logratio})  with a passage to an equivalent structure by translation of
$\log \frac{S_T/ES_T}{S_{t_0}/ES_{t_0}}$ or a change of the normal probability conditional on the value of the Poissonian component. The translation's amount
is determined such that (\ref{eq:weneed}) holds and has the value
\begin{equation}
\label{eq:translALL}
r(T-t_0)-\log \frac{ES_T}{ES_{t_0}}- \mu_{[t_0,T]}-\frac{\sigma^2_{[t_0,T]} }{2}-\log M_{L^*_T}(1),
\end{equation}
with $ M_{L^*_T}(t)$ the  moment generating function of the Poissonian component evaluated at $t=1.$

\beth \label{t:infdiv}
When ${\cal A}1-{\cal A}3$ hold, from Theorem  \ref{t:LLCmain}
it follows that
\begin{equation}
(i) \  \lim_{n\rightarrow \infty}P_{t_0,n}[\log \Pi_{j=1}^{k_n}
\frac{p_{t_{j}^n}}{p_{t_{j-1}^n}}\le x]=
\int \Phi(\frac{x-y-\mu_{[t_0,T]}}
{\sigma_{[t_0,T]}}) L^*_{t_0}(dy),
\label{eq:levyprob1}
\end{equation}
\begin{equation}
(ii) \ \lim_{n\rightarrow \infty}P_{T,n}[\log \Pi_{j=1}^{k_n}
\frac{p_{t_{j}^n}}{p_{t_{j-1}^n}}\le x]=\int
 \Phi(\frac{x-y+\mu_{[t_0,T]}}
{\sigma_{[t_0,T]}}) L^*_T(dy).
\label{eq:levyprob2}
\end{equation}

\enth

\bec \label{c:ommitance}
Under the assumptions of Theorem  \ref{t:infdiv}, (\ref{eq:levyprob1})
and (\ref{eq:levyprob2}) both hold when one of the terms in the product
$\Pi_{j=1}^{k_n}
\frac{p_{t_{j}^n}}{p_{t_{j-1}^n}}$ is ommited.
\enc

\ber
Without ${\cal A}3$ 
 the distribution of $\log \frac{S_T}{S_{t_0}}$ is infinitely divisible
when the 
random variables $\{Y_{n,j}, \ j=1,\ldots, k_n\}$ and
$\{U_{n,j}, \ j=1,\ldots, k_n\}$ are uniformly asymptotically negligible.
However, computational difficulties arise because centering is
needed at truncated expectations that are also used in the definition
of the sequences of L\'{e}vy measures $L_{k_n}$ in 
Theorem  \ref{t:LLCmain}.
The limit law is more
complicated than the one presented herein (Lo\`{e}ve (1977)).
\enr


The next result confirms the martingale property of the prices-densities.

\beth 
\label{t:MG}
When the ratios of prices-densities $\{\frac{p_{t_j^n}}{p_{t_{j-1}^n}}, 
\ j=1,\ldots, k_n\}$ are independent of $S_{t_0},$ the price-densities
are a martingale under $P_{t_0,n}.$ 
\enth

\subsection{Calm Stock}


\quad Calm stock (or calm share price) has 
prices-densities $p_{t+\delta}, \ p_t$ (see (\ref{eq:pricedensities}))
that
do not differ much {\it often} with respect 
to $P$ 
for small $\delta$-values, thus excluding the case of unusual jumps.
To provide a quantitative definition of  calm share, 
the difference of $p_{t+\delta}$ and $p_t$ 
over the (forward) regions 
\begin{equation}
\{|\frac{\sqrt{p_{t+\delta}}}{\sqrt{p_t}}-1|>\epsilon\},  \ \epsilon>0,
\label{eq:forwardregion}
\end{equation}
is measured by 
\[
\int(\sqrt{p_{t+\delta}}-\sqrt{p_t})^2
I(|\frac{\sqrt{p_{t+\delta}}}{\sqrt{p_t}}-1|>\epsilon)dP;
\]
$I$
is the
indicator function.  

\bef
Let $t_1^n<\ldots< t^n_{k_n-1}$ be a partition  of $(t_0=t_0^n, T=t^n_{k_n}),$
with mesh size $\delta_n=\sup \{t^n_j-t^n_{j-1}, \ j=1,\ldots, k_n\}$ and
$\epsilon>0.$
The stock $\{S_t\}$ is $\epsilon$-calm in $[t_0,T]$ if for any partition
\begin{equation}
\lim_{\delta_n \rightarrow 0 } 
\sum_{j=1}^{k_n} \int (\sqrt{p_{t_j^n}}-\sqrt{p_{t^n_{j-1}}})^2
I(|\sqrt{
\frac{p_{t_j^n}}{p_{t^n_{j-1}}}}-1|>\epsilon) dP=0.
\label{eq:ecalmexp}
\end{equation}
\enf
\bef
The stock $\{S_t\}$ is calm in $[t_0,T]$ if it is $\epsilon$-calm for every 
$\epsilon>0.$
\enf

For calm stock, 
the random variables $Y_{n,j}, \ j=1,\ldots,k_n$ (see (\ref{eq:yu}))
satisfy Lindeberg's condition since
(\ref{eq:ecalmexp}) can be written 
\begin{equation}
\lim_{k_n \rightarrow \infty}\sum_{j=1}^{k_n} E_{P_{t_{j-1}^n}}Y_{n,j}^2
I(|Y_{n,j}|>\epsilon)=0.
\label{eq:eqecalmexp}
\end{equation}
When $Y_{n,j},  \ j=1,\ldots, k_n,$ are 
independent and uniformly asymptotically negligible,
Lindeberg's condition is necessary and sufficient
condition for $\sum_{j=1}^{k_n}Y_{n,j}$ to have
asymptotically a normal distribution and is thus
satisfied for the prices-densities of geometric
Brownian motion. 

\ber
From (\ref{eq:edcalm}) and (\ref{eq:varnormcomp}) 
it follows that (\ref{eq:eqecalmexp}) 
and all its implications (including option pricing in section 4) hold also for
$U_{n,j}, \ j=1,\ldots,k_n,$ and that in (\ref{eq:ecalmexp}) instead of the indicators
of the forward region (\ref{eq:forwardregion})
one can use indicators of backward regions 
$I(|\sqrt {\frac{ p_{t_{j-1}^n} }{p_{t^n_j}} }-1|>\epsilon), \ j=1,\ldots, k_n,$
i.e. we can  define forward and backward calm stock and the obtained results hold for both.
\enr




\subsection{\bf Calm stock-Normal distribution for $\log \frac{S_T}{S_{t_0}}$
}

The distribution of $\log \frac{S_T/ES_T}{S_{t_0}/ES_{t_0}}$ is initially obtained.

\beth \label{t:asydicalm}
When ${\bf {\cal A}}1, {\bf {\cal A}}2, \mbox{ and }{\bf {\cal A}}3 (ii)$
hold for a calm stock in  $[0,T],$
there is $\sigma_{[t_0,T]}>0$ such that\\
(i) \hspace{2ex}$\{P_{t_0,n}\}$ and  $\{P_{T,n}\}$ are contiguous, i.e.
$\lim_{n \rightarrow \infty}P_{t_0,n}(A_n)=0 \Longleftrightarrow 
\lim_{n \rightarrow \infty} P_{T,n}(A_n)=0,$
\begin{equation}
(ii)
\hspace{2ex}
lim_{n\rightarrow \infty}P_{t_0,n}
[\log \Pi_{j=1}^{k_{n}}
\frac{p_{t_j^{n}}}{p_{t^{n}_{j-1}}}\le x]=
\Phi(\frac{x+\frac{\sigma^2_{[t_0,T]}}{2}} {\sigma_{[t_0,T]}}),
\label{eq:traderprob}
\end{equation}
\begin{equation}
(iii) \hspace{2ex} lim_{n\rightarrow \infty}P_{T,n}[\log \Pi_{j=1}^{k_{n}}
\frac{p_{t_j^{n}}}{p_{t^{n}_{j-1}}}\le x]=
\Phi(\frac{x-\frac{ \sigma^2_{[t_0,T]}}{2}} {\sigma_{[t_0,T]}}),
\label{eq:buyerprob}
\end{equation}
\begin{equation}
(iv) \hspace{2ex}
\sigma^2_{[t_0,T]}=4\lim_{\tau \rightarrow 0} \lim_{n \rightarrow \infty}
\sum_{j=1}^{k_{n}}E_{P_{t^n_{j-1}}}
(\sqrt{ \frac{ p_{t_j^n} } { p_{t^n_{j-1}} } } -1)^2
I(|\sqrt{ \frac{ p_{t_j^n} } { p_{t^n_{j-1}} } } -1| \le \tau).
\label{eq:variancelambda}
\end{equation}
$\hfill \Box$

\enth

The risk neutral probability
 is obtained from Theorem \ref{t:asydicalm}
using a passage to an equivalent structure with translation
of $\log \frac{S_T/ES_T}{S_{t_0}/ES_{t_0}}$  by $r(T-t_0)- \log \frac{ES_T}{ES_{t_0}}$ such that 
(\ref{eq:weneed}) holds. Note that for calm stock the last three terms of translation (\ref{eq:translALL}) vanish.

The result indicating clearly volatility's role follows.

\bec \label{c:convexp}
Under the assumptions  of Theorem \ref{t:asydicalm}, from
 (\ref{eq:traderprob}) and (\ref{eq:buyerprob})  it follows 
that  the binary experiment
${\cal E}_{k_n}=\{P_{t_0,n}, P_{T,n}\}$ 
converges  
to the Gaussian experiment
${\cal G}=\{P_0=N(0,1), \ P_T=N(\sigma_{[t_0,T]},1)\}$
when $k_n \rightarrow \infty.$ 
From the form of ${\cal G}$ it is
clear that the volatility, $\sigma_{[t_0,T]},$
is the determining factor in the buyer's
decision.
\enc

\ber The parameter $\sigma^2_{[t_0,T]}$ does not necessarily have the form
$\sigma^2(T-t_0)$ obtained, for example, with the
B-S-M model. 
Conditions are  provided for  ${\cal A} 3 $ to hold  and for 
$\sigma^2_{[t_0,T]}$ to have the form $\sigma^2 (T-t_0), \ \sigma>0$ (Proposition  \ref{p:T-t}).
\enr

\section{Pricing the share's European call option}



\subsection{Discounting with expectations' ratios and interest}

\quad  We already used  $a^{-1}{[t,T]}$ (see (\ref {eq:a}))
to discount the stock price $S_T$ at $t$ and
obtain the distribution of $\log \frac{S_T}{S_{t}}.$  
We will use 
it also to price the stock's portfolio. This is not restrictive since, when calculating expected values under the
martingale probability obtained from the equivalent structure,  $a^{-1}{[t,T]}$ coincides with $e^{-r(T-t)}.$


 The price of the call option
at
$t=t_0$ is
\begin{equation}
E_{\tilde Q_n} a^{-1}[t_0,T] S_T I(S_T>X)-Xe^{-r(T-t_0)}E_{\tilde Q_n}I(S_T>X);
\label{eq:expectedcostatT}
\end{equation}
the expectation $E_{\tilde Q_n}$ is calculated with respect to probability
$\tilde Q_n=P_{t_0,n}, \ P_{T, n}.$

To calculate the expectations for the trader and the buyer  
in (\ref{eq:expectedcostatT}) 
via the distribution previously obtained
for $\log \frac{S_T}{S_{t_0}}$ 
rewrite
(\ref{eq:expectedcostatT}) using prices-densities (\ref{eq:pricedensities})
at the deterministic transaction times to obtain
\begin{equation}
E_{\tilde Q_n} M_{k_n} S_{t_0} I(M_{k_n}>\frac{X}{S_{t_0}}a^{-1}[t_0,T]) 
-Xe^{-r(T-t_0)}E_{\tilde Q_n}I(M_{k_n}>\frac{X}{S_{t_0}}a^{-1}[t_0,T]);
\label{eq:expectedcostatTwithprobdenProduct}
\end{equation}
the term 
\begin{equation}
M_{k_n}=\frac{S_T}{S_{t_0}}a^{-1}[t_0,T]=\frac{\Pi_{j=1}^{k_{n}}
p_{t_j^{n}} }
{\Pi_{j=1}^{k_{n}}
p_{t_{j-1}^n} }.
\label{eq:likelihoodratiom}
\end{equation}

Note that the price can be obtained using either {\bf 1)} or {\bf 2)}:\\
{\bf 1)} {\bf a)} discount  the first term in (\ref{eq:expectedcostatT}) with 
$a^{-1}{[t_0,T]},$ \\  {\bf b)}  take the limit of the expectations with respect to $P_{t_0,n},$ \\
{\bf c)} pass by translation to the risk neutral equivalent structure and ``fair''  price.\\
{\bf 2)}  {\bf  a)} discount  the first term in (\ref{eq:expectedcostatT}) with $e^{-r(T-t_0)},$\\
{\bf b)} translate  ``properly''  $P_{t_0,n}$   to obtain $P_{t_0,n}^*$ without the term involving $a{[t_0,T]},$  \\  
{\bf c)} take the limit of the expectations  to obtain the  ``fair''  price.

We preferred to use {\bf 1)} obtaining (\ref{eq:expectedcostatTwithprobdenProduct})  for easier exposition,  because the translation {\bf 2 b)} for non-calm shares is conditional on 
the value of the Poissonian component and is not easily described. We also preferred not to use directly the limit  risk neutral probability when taking the expected value of the interest discounted call's payoff.

Two different situations are considered for pricing the option at $t_0:$

$({\cal P}1)$ At $t<t_0,$
the trader prices the option at $t_0$  given that
$S_{t_0}=s_{t_0}.$ 
Then, the expectations in 
(\ref{eq:expectedcostatTwithprobdenProduct})
are conditional on $S_{t_0}=s_{t_0}.$

$({\cal P}2)$ To price the option at $t_0$
when the value $S_{t_0}=s_{t_0}$ is known,
the trader calculates the option's price at
$t=t_1^n>t_0$ with $t_1^n$ decreasing to $t_0$ with $n.$ 
In (\ref{eq:expectedcostatTwithprobdenProduct}) and (\ref{eq:likelihoodratiom})
$t_0$ is replaced by $t_1^n.$  

In both $({\cal P}1)$ and $({\cal P}2)$
we let in (\ref{eq:expectedcostatTwithprobdenProduct}) 
$n$ and/or the number of transactions $k_n$  
increase to infinity to obtain at $t_0$ the trader's option price 
and a lower bound on the buyer's price, that agree
under $({\cal P}1)$ and $({\cal P}2).$ A translation provides the risk-neutral price and lower bound.


\subsection{The Pricing Assumptions}


\quad
In addition to the modeling assumptions ${({\cal A}1)}-{({\cal A}3)}$
assume

${({\cal A}4)}$ The market 
consists of the stock $S$ and a riskless bond
that appreciates at fixed rate $r$ and there are no dividends or
transaction costs. The option is European. The buyer prefers to 
pay less than more. 

For $({\cal P}1)$ pricing assume

${({\cal A}5)}$ The ratios
 $\frac{p_{t_j^n}}
{p_{t_{j-1}^n}}, j=1,\ldots, k_n$ are independent of $S_{t_0}.$

For $({\cal P}2)$ pricing assume

${({\cal A}5^*)}$ At $t_0,$
$\lim_{n \rightarrow \infty} S_{t_1^n}=s_{t_0}$ in probability,
$\lim_{n \rightarrow \infty} ES_{t_1^n}=s_{t_0}.$ 

In the  pricing sections, to obtain the {\it B-S-M} price $\log c$ denotes  $\ln c.$

\subsection{Pricing a European option using ${({\cal A}5)}$}

\quad Pricing under $({\cal P}1)$ follows. The components in  (\ref{eq:expectedcostatTwithprobdenProduct}) are initially calculated.

\beth \label{t:pri1terms}
When ${\cal A}1-{\cal A}5$ hold, 
the limits of the terms in (\ref{eq:expectedcostatTwithprobdenProduct})
with $\tilde Q_n=P_{t_0,n}, \ P_{T,n},$ and $M_{k_n}$ as in (\ref{eq:likelihoodratiom}),
are: 
\[
(i) \ \lim_{n\rightarrow\infty}P_{t_0,n}[M_{k_n}>\frac{X}{S_{t_0}}a^{-1}[t_0,T]
|S_{t_0}=s_{t_0}]
\]
\[
=\int \Phi(\frac { \log(s_{t_0}/X)+\log a[t_0,T]+\mu_{[t_0,T]}+y}
{\sigma_{[t_0,T]}})L^*_{t_0}(dy),
\]
$$
(ii) \lim_{n\rightarrow\infty}E_{P_{t_0,n}}[S_{t_0}M_{k_n}
I(M_{k_n}>\frac{X}{S_{t_0}}a^{-1}[t_0,T])|S_{t_0}=s_{t_0}]
=s_{t_0}\lim_{n\rightarrow\infty}P_{T,n}[M_{k_n}>\frac{X}{S_{t_0}}a^{-1}[t_0,T]
]$$
\[
=s_{t_0}\int \Phi(\frac { \log(s_{t_0}/X)+\log a[t_0,T]
-\mu_{[t_0,T]}+y}{\sigma_{[t_0,T]}})L^*_T(dy),
\]
$$
(iii) \  \liminf E_{P_{T,n}}[S_{t_0}M_{k_n}
I(M_{k_n}>\frac{X}{S_{t_0}}a^{-1}[t_0,T])|S_{t_0}=s_{t_0}]
$$
\[
\ge s_{t_0}e^{-\mu_{[t_0,T]}+.5\sigma^2_{[t_0,T]}}
\int e^y 
\Phi(\frac{\log(s_{t_0}a[t_0,T]/X)-\mu_{[t_0,T]}+\sigma^2_{[t_0,T]}
+y}{\sigma_{[t_0,T]}}) L^*_T(dy).
\]
\enth


Let 
\begin{equation}
Q_{t_0,n}(x)=P_{t_0,n}[\log \Pi_{j=1}^{k_{n}} 
\frac { p_{t_j^{n}} }{ p_{t^{n}_{j-1}} }\le x], \hspace{3ex}
Q_{T,n}(x)=P_{T,n}[\log \Pi_{j=1}^{k_{n}} 
\frac { p_{t_j^{n}} }{ p_{t^{n}_{j-1}} }\le x],
\label{eq:probR}
\end{equation}
be cumulative distributions on the real line.
The option's price is obtained below and
in particular, 
by passing to an equivalent structure,
a generalization of the {\it B-S-M} price 
similar to the price in Merton (1976).

\bec \label{c:pri1}
Under $({\cal P}1)$ pricing, 
when 
${\cal A}1-{\cal A}5$ hold: \\
(i) the trader's 
price for the European option is
$$
C(t_0)=
s_{t_0}\int \Phi(\frac { \log(s_{t_0}/X)+
\log a[t_0,T]
-\mu_{[t_0,T]}+y}{\sigma_{[t_0,T]}})L^*_T(dy)$$
\[
-Xe^{-r(T-{t_0})}
\int \Phi(\frac { \log(s_{t_0}/X)+\log a[t_0,T]+\mu_{[t_0,T]}+y}
{\sigma_{[t_0,T]}})L^*_{t_0}(dy),
\]
(ii) a lower bound on the buyer's price is
$$
s_{t_0}e^{-\mu_{[t_0,T]}+.5\sigma^2_{[t_0,T]}}
\int e^y
\Phi(\frac{\log(s_{t_0}/X)+
\log a[t_0,T]
-\mu_{[t_0,T]}+\sigma^2_{[t_0,T]}
+y}{\sigma_{[t_0,T]}}) L^*_T(dy)$$
\[
-Xe^{-r(T-{t_0})}
\int\Phi(\frac { \log(s_{t_0}/X)+\log a[t_0,T]
-\mu_{[t_0,T]}+y}{\sigma_{[t_0,T]}})L^*_T(dy).
\]
(iii) By passing to an equivalent structure
either by translation (\ref{eq:translALL}) or by change of the normal probability
conditional on the Poissonian component $\log a[t_0,T]$ is replaced
in i) and  ii)  via  
(\ref{eq:translALL}) by
\begin{equation}
\label{eq:replpga}
r(T-t_0)- \mu_{[t_0,T]}-\frac{\sigma^2_{[t_0,T]} }{2}-\log M_{L^*_T}(1),
\end{equation}
such that (\ref{eq:weneed}) holds and the ``fair'' price is obtained.
\enc

The results for calm stock follow.

\beth \label{t:pri1termscalm} For calm stock, under  assumptions
${\cal A}1, {\cal A}2, {\cal A}3 (ii), {\cal A}4, {\cal A}5:$
\[
(i) \ {\lim}_{n \rightarrow \infty} P_{t_0,n}[M_{k_n}>\frac{X}{S_{t_0}}
a^{-1}[t_0,T]|S_{t_0}=s_{t_0}]=
\Phi(\frac{\log (s_{t_0}/X)+\log a[t_0,T]-\frac{\sigma^2_{[t_0,T]}}{2}}
{\sigma_{[t_0,T]}}),
\]
(ii)
\[{\lim}_{n \rightarrow \infty} E_{P_{t_0,{n}}}
[S_{t_0}M_{k_n}I(M_{k_n}>\frac{X}{S_{t_0}}a^{-1}[t_0,T])|S_{t_0}=s_{t_0}]
= s_{t_0}
{\lim}_{n \rightarrow \infty} P_{T,n}[M_{k_n}>\frac{X}{s_{t_0}}a^{-1}[t_0,T]]
\]
\[
=s_{t_0} 
\Phi(\frac{\log (s_{t_0}/X)+\log a[t_0,T]+\frac{\sigma^2_{[t_0,T]}}{2}}
{\sigma_{[t_0,T]}}),
\]
\[
(iii) \liminf E_{P_{T,{n}}}
[S_{t_0}M_{k_n}I(M_{k_n}>\frac{X}{S_{t_0}}a^{-1}[t_0,T])|S_{t_0}=s_{t_0}]
\]
\[
\ge s_{t_0} e^{\sigma^2_{[t_0,T]}}
\Phi(\frac{\log (s_{t_0}/X)+\log a[t_0,T]+1.5\sigma^2_{[t_0,T]}}
{\sigma_{[t_0,T]}}).
\]
\enth

Comparing {\em i)} and  {\em ii)} it follows that the probability $S_T$ is greater than $X$ is larger for the buyer than for the trader.
The same holds for non-calm stock from Theorem 4.1  {\em i)},  {\em ii)} since $\mu_{[t_0,T]}$ is less than zero.

\bec \label{c:pri1calm}
For calm stock, under $({\cal P}1)$ pricing and
${\cal A}1, {\cal A}2, {\cal A}3 (ii), {\cal A}4, {\cal A}5:$\\
(i) the trader's price for the European option
is 
\begin{equation}
\label{eq:bsmspr}
C(t_0)=s_{t_0} \Phi(d_1)-Xe^{-r(T-{t_0})}  \Phi(d_2)
\end{equation}
$$=s_{t_0} \lim_{n \rightarrow \infty} P_{T,n}(S_T>X)-Xe^{-r(T-t_0)}
\lim_{n \rightarrow \infty} P_{t_0,n}(S_T>X),$$
with
$$d_1=\frac{\log (s_{t_0}/X)+\log a[t_0,T]+\frac{\sigma^2_{[t_0,T]}}{2}}
{\sigma_{[t_0,T]}}, \mbox{   } d_2=d_1-\sigma_{[t_0,T]},$$
(ii) a lower bound on the buyer's price is
$$s_{t_0}e^{\sigma^2_{[t_0,T]}}\Phi(\tilde d_1)-
Xe^{-r(T-{t_0})}\Phi(\tilde d_2),$$
with
$$\tilde d_1=\frac{\log (s_{t_0}/X)+\log a[t_0,T]+1.5\sigma^2_{[t_0,T]}}
{\sigma_{[t_0,T]}}, \ \tilde d_2=\tilde d_1-\sigma_{[t_0,T]}.$$
(iii) By translating either $Q_{t_0,n}, \ Q_{T,n}$ in (\ref{eq:probR}) or the limit normal distributions
 by $r(T-t_0)-\log a_{[t_0,T]}$ or with a change of probability via Lemma 5.4
in (i) (resp. (ii) ) 
$\log a[t_0,T]$ is replaced by $r(T-t_0)$ in $d_1$ (resp. $\tilde d_1$),
thus obtaining the {\it B-S-M} price (resp. a 
lower bound on the associated buyer's
price).
\enc

From {\em ii)} it follows that buyer's price includes a volatility premium.

\ber \label{r:stochint1}
For stochastic interest rate $R,$ with ``expected accumulation'' function
$a_R[t_0,T]$ similar results can be obtained as in Corollaries  
\ref{c:pri1} and 
 \ref{c:pri1calm}: 
in {\it (i), (ii)} with $a^{-1}_R[t_0,T]$ replacing $e^{-r(T-t_0)},$
and in {\it (iii)} 
by considering the equivalent structure 
to replace $\log a[t_0,T]$ by $\log a_R[t_0,T].$
\enr

\subsection{Pricing a European option using ${({\cal A}5^*)}$}

\quad We replace $t_0$ in the previous sections by $t_1^n$ and use $P_{t_1^n,n}$
for the trader's strategy and $P_{T,n}$ (abuse of notation) for
the buyer's strategy. $(\Omega, {\cal F}, {\cal P})$ is the probability
space at $t_0, \ S_{t_0}=s_{t_0} \ a.s. \ {\cal P}.$
The asymptotic distributions previously obtained remain because
of the asymptotic negligibility assumption ${\cal A}3(i)$ and
Corollary 4.2. $ES_{t_0}$ is replaced by $s_{t_0}.$
Instead of (\ref{eq:expectedcostatTwithprobdenProduct}) and 
(\ref{eq:likelihoodratiom})
we use  
$$\tilde M_{k_n}=\frac{S_T}{S_{t_1^n}}a^{-1}[t_1^n,T]=
\frac{ \Pi_{j=2}^{k_{n}}
p_{t_j^{n}} }
{ \Pi_{j=2}^{k_{n}}p_{t_{j-1}^n} },$$
\begin{equation}
E_{Q_n} \tilde M_{k_n} S_{t_1^n} I(\tilde M_{k_n}>\frac{X}{S_{t_1^n}}a^{-1}[t_1^n,T])
-Xe^{-r(T-t_1^n)}E_QI(\tilde M_{k_n}>\frac{X}{S_{t_1^n}}a^{-1}[t_1^n,T]).
\label{eq:expectedcostatTwithprobdenProductreduced}
\end{equation}
In this section $a[t_0,T]=ES_T/s_{t_0}$ but $a[t,T]=ES_T/ES_t, \ t>t_0.$

\beth \label{t:pri2terms}
When ${\cal A}1-{\cal A}5^*$ hold, 
the limits of the terms in (\ref{eq:expectedcostatTwithprobdenProductreduced}),
with $\tilde Q_n=P_{t_1^n,n}, \ P_{T,n},$ are: 
\[
(i) \ \lim_{n\rightarrow\infty}P_{t_1^n,n}[\tilde M_{k_n}>\frac{X}{
S_{t_1^n}}a^{-1}[t_1^n,T]
]
=\int \Phi(\frac { \log(s_{t_0}/X)+\log a[t_0,T]+\mu_{[t_0,T]}+y}
{\sigma_{[t_0,T]}})L^*_{t_0}(dy),
\]
$$
(ii) \lim_{n\rightarrow\infty}E_{P_{t_1^n,n}}[S_{t_1^n}\tilde M_{k_n}
I(\tilde M_{k_n}>\frac{X}{S_{t_1^n}}a^{-1}[t_1^n,T])]
=s_{t_0}\lim_{n\rightarrow\infty}P_{T,n}[\tilde M_{k_n}>\frac{X}{S_{t_1^n}}
a^{-1}[t_1^n,T]
]$$
\[
=s_{t_0}\int \Phi(\frac { \log(s_{t_0}/X)+\log a[t_0,T]
-\mu_{[t_0,T]}+y}{\sigma_{[t_0,T]}})L^*_T(dy),
\]
$$
(iii) \  \liminf E_{P_{T,n}}[S_{t_1^n} \tilde M_{k_n}
I(\tilde M_{k_n}>\frac{X}{S_{t_1^n}}a^{-1}[t_1^n,T])]
$$
\[
\ge s_{t_0}e^{-\mu_{[t_0,T]}+.5\sigma^2_{[t_0,T]}}
\int e^y 
\Phi(\frac{\log(s_{t_0}a[t_0,T]/X)-\mu_{[t_0,T]}+\sigma^2_{[t_0,T]}
+y}{\sigma_{[t_0,T]}}) L^*_T(dy).
\]
\enth


Let
$$ 
Q_{t_1^n,n}(x)=P_{t_1^n,n}[\log \Pi_{j=2}^{k_{n}} 
\frac { p_{t_j^{n}} }{ p_{t^{n}_{j-1}} }\le x], \hspace{3ex}
Q_{T,n}(x)=P_{T,n}[\log \Pi_{j=2}^{k_{n}} 
\frac { p_{t_j^{n}} }{ p_{t^{n}_{j-1}} }\le x],
$$
be cumulative distributions on the real line.
The option's price is obtained below
and in particular, 
by passing to an equivalent structure as in the previous section, a generalization of the {\it B-S-M} price 
similar to the price in Merton (1976).

\bec \label{c:pri2}
Under $({\cal P}2)$ pricing, 
when 
${\cal A}1-{\cal A}5^*$ hold
the obtained prices in Corollary  \ref{c:pri1} remain valid.
\enc

\beth \label{t:pri2termscalm}
For calm stock, under  assumptions
${\cal A}1, {\cal A}2, {\cal A}3 (ii), {\cal A}4, {\cal A}5^*:$
\[
(i) \ {\lim}_{n \rightarrow \infty} P_{t_1^n,n}[\tilde M_{k_n}>\frac{X}{S_{t_r^n}}
a^{-1}[t_1^n,T]]=
\Phi(\frac{\log (s_{t_0}/X)+\log a[t_0,T]-\frac{\sigma^2_{[t_0,T]}}{2}}
{\sigma_{[t_0,T]}}),
\]
(ii)
\[{\lim}_{n \rightarrow \infty} E_{P_{t_1^n,{n}}}
[S_{t_1^n}\tilde M_{k_n}I(\tilde M_{k_n}>\frac{X}{S_{t_1^n}}a^{-1}[t_1^n,T])]
= s_{t_0}
{\lim}_{n \rightarrow \infty} 
P_{T,n}[\tilde M_{k_n}>\frac{X}{S_{t_1^n}}a^{-1}[t_1^n,T]]
\]
\[
=s_{t_0} 
\Phi(\frac{\log (s_{t_0}/X)+\log a[t_0,T]+\frac{\sigma^2_{[t_0,T]}}{2}}
{\sigma_{[t_0,T]}}),
\]
\[
(iii) \liminf E_{P_{T,{n}}}
[S_{t_1^n}\tilde M_{k_n}I(\tilde M_{k_n}>\frac{X}{S_{t_1^n}}a^{-1}[t_1^n,T])]
\]
\[
\ge s_{t_0} e^{\sigma^2_{[t_0,T]}}
\Phi(\frac{\log (s_{t_0}/X)+\log a[t_0,T]+1.5\sigma^2_{[t_0,T]}}
{\sigma_{[t_0,T]}}).
\]
\enth

\bec \label{c:pri2calm} For calm stock, under $({\cal P}2)$ pricing and
${\cal A}1, {\cal A}2, {\cal A}3 (ii), {\cal A}4, {\cal A}5^*$
the obtained prices in Corollary  \ref{c:pri1calm} remain valid.
\enc

\ber \label{r:stochint2} 
The results remain valid with
stochastic interest rate $R$ as described in Remark  \ref{r:stochint1}.
\enr



\section{Appendix}

\subsection{Proofs}


{\bf Proof of Proposition  \ref{p:Yaugm}- 
Independence of $\tilde Y_{n,j}, \tilde U_{n,j}, 
\frac{p_{t_j^n}}{p_{t_{j-1}^n}}, 
\ j=1,\ldots, k_n$ under $P_{t_0,n}, P_{T,n}$}

Define on $(\Omega^{k_n},
{\cal F}^{k_n})$ random variables
$$\tilde Y_{n,j}(\omega^{(k_n)})=Y_{n,j}(\omega_j), \ j=1,\ldots, k_n,$$
$\omega^{(k_n)}=(\omega_1,\ldots, \omega_{k_n}).$
Then, 
$\tilde Y_{n,j}, \ j=1,\ldots,k_n,$ are independent with respect to $P_{t_0,n}$ 
(defined in (\ref{eq:prodmeas})),
i.e. 
\begin{equation}
P_{t_0,n}[\cap_{j=1}^{k_n}\{ \tilde Y_{n,j} \in B_j\}]=\Pi_{j=1}^{k_n}
P_{t_0,n}[\tilde Y_{n,j} \in B_j].
\label{eq:indep}
\end{equation}
Equality (\ref{eq:indep}) follows from (\ref{eq:prodmeas}) and equality
$$ \cap_{j=1}^{k_n}\{\omega^{(k_n)}:\tilde Y_{n,j}(\omega^{(k_n)}) \in B_j\}
=\{\omega_1:Y_{n,1}(\omega_1) \in B_1\}x\ldots x
\{\omega_{k_n}:Y_{n,k_n}(\omega_{k_n}) \in B_{k_n}\}
$$
which holds since
\[
\{\omega^{(k_n)}:\tilde Y_{n,j}(\omega^{(k_n)}) 
\in B_j\}=\Omega x \ldots x \{\omega_j:
Y_{n,j}(\omega_j) \in B_j\} x \ldots x \Omega, \ j=1,\ldots, k_n.
\]
One can confirm in the same way
independence of $\tilde Y_{n,j}, j=1,\ldots, k_n,$
with respect to $P_{T,n}.$ 
The same results hold for 
$\tilde U_{n,j}, \ j=1,\ldots,k_n$ defined similarly on 
$(\Omega^{k_n}, {\cal F}^{k_n})$ from 
$U_{n,j}, \ j=1,\ldots,k_n.$
Since 
\[
(Y_{n,j}+1)^2=\frac{p_{t,j}^n}{p_{t,j-1}^n}, \ j=1,\ldots, k_n,
\]
independence, in the same sense, of the price-densities ratios 
follows. $\hspace{3ex} \hfill \Box$

\bep \label{p:YUequiv} 
(i) For $Y_{n,j}, \ U_{n,j}, \ j=1,\ldots, k_n,$ in
(\ref{eq:yu}) and for $\epsilon$ small,
\begin{equation}
\sum_{j=1}^{k_n}E_{P_{t_{j-1}^n}}Y_{n,j}^2I(|Y_{n,j}|>\frac{\epsilon}{1-2\epsilon})
\le \sum_{j=1}^{k_n}E_{P_{t_j^n}}U_{n,j}^2I(|U_{n,j}|>
\frac{\epsilon}{1-\epsilon})\le
\sum_{j=1}^{k_n}E_{P_{t_{j-1}^n}}Y_{n,j}^2I(|Y_{n,j}|>\epsilon).
\label{eq:edcalm}
\end{equation}
(ii) If \ $\sum_{j=1}^{k_n}E_{P_{t_{j-1}^n}}Y_{n,j}^2<+\infty,$ then
\begin{equation}
\lim_{\tau \rightarrow 0} \lim_{n \rightarrow \infty} \sum_{j=1}^{k_n}
E_{P_{t_{j-1}^n}}Y_{n,j}^2I(|Y_{n,j}|\le \tau)
=\lim_{\tau \rightarrow 0} \lim_{n \rightarrow \infty} \sum_{j=1}^{k_n}
E_{P_{t_j^n}}U_{n,j}^2I(|U_{n,j}|\le \tau).
\label{eq:varnormcomp}
\end{equation}
\enp

\ber \label{r:YUequiv} Proposition \ref{p:YUequiv} shows that if  
$\{Y_{n,j}, \ j=1,\ldots,k_n\}$ satisfy Lindeberg's condition the same 
holds for $\{U_{n,j}, \ j=1,\ldots,k_n\},$ and conversely.
\enr

{\bf Proof of Proposition \ref{p:YUequiv}}

(i)

Note that for densities $f$ and $g$ it holds
\[
I(|\sqrt{\frac{f}{g}}-1|>\epsilon)=1 \Longleftrightarrow
\{\sqrt{\frac{f}{g}}-1>\epsilon\} \mbox{  or  }
\{\sqrt{\frac{f}{g}}-1 < -\epsilon\}.
\]
and that for small $\epsilon$
it holds
$$\{|\sqrt{\frac{f}{g}}-1|>\epsilon\}=
\{\sqrt{\frac{g}{f}}<1-\frac{\epsilon}{1+\epsilon}\}
\cup \{\sqrt{\frac{g}{f}}>1+\frac{\epsilon}{1-\epsilon}\}
\supseteq \{|\sqrt{\frac{g}{f}}-1|>\frac{\epsilon}{1-\epsilon}\}.$$
(ii) It follows from (i) since $\sum_{j=1}^{k_n}E_{P_{t_{j-1}^n}}Y_{n,j}^2
=\sum_{j=1}^{k_n}E_{P_{t_j^n}}U_{n,j}^2.$
$\hspace{3ex} \hfill \Box$

{\bf Proof of Corollary  \ref{c:lambda}} Follows from 
(\ref{eq:varnormcomp}),
the equalities $E_{P_{t_{j-1}^n}}Y_{n,j}=
E_{P_{t_j^n}}U_{n,j}, \ j=1,\ldots, k_n,$ and the definition
of 
$L_{t_0}$ (in Theorem  \ref{t:LLCmain}). $\hspace{3ex} \hfill \Box$

{\bf Proof of Theorem  \ref{t:infdiv}-Infinitely divisible distribution
of $\log \frac{S_T}{S_{t_0}}$} 

From Theorem \ref{t:LLCmain}, 
the distribution of $\Lambda_{t_0}$ (resp. $\Lambda_T$)
is that of a sum of  a normal random variable $X$ and  
a Poissonian random variable $Y$ that are independent.
Parts $(i)$ and $(ii)$ follow
by conditioning the limit
distribution on the value $Y=y$ of the Poissonian component,
and then integrating with $L^*_{t_0}$ and $L^*_T$ respectively. $\hspace{3ex} \hfill \Box$



{\bf Proof of Theorem \ref{t:MG}-The martingale property of prices-densities} 

It is enough to show that
$$ E_{P_{t_0,n}}
[p_{t^n_{j+k}}|p_{t_j^n}, p_{t^n_{j-1}}, \ldots, p_{t_1^n},
p_{t_0}]=p_{t_j^n}.$$
Observe that
\begin{equation}
E_{P_{t^n_{j-1}}} \frac{p_{t_j^n}}{p_{t^n_{j-1}}}=
\int \frac{p_{t_j^n}}{p_{t^n_{j-1}}} p_{t^n_{j-1}}dP
=\int\frac{S_{t_j^n}}{ES_{t_j^n}}dP=1, \ j=1,\ldots, k_n.
\label{eq:prdeprop}
\end{equation}

From the independence of prices-densities ratios
and (\ref{eq:prdeprop})
$$E_{P_{t_0,n}}
[\frac{p_{t^n_{j+k}}}{p_{t_j^n}}|p_{t_j^n}, p_{t^n_{j-1}},
\ldots, p_{t_1^n},
p_{t_0}]=
E_{P_{t_0,n}}
[\Pi_{m=1}^k \frac{p_{t^n_{j+m}}}{p^n_{t_{j+m-1}}}
|\frac{p_{t_j^n}}{p_{p^n_{j-1}}},\ldots, \frac{p_{t_1^n}}{p_{t_0}},
p_{t_0}]$$
$$=\Pi_{m=1}^kE_{P_{t^n_{j+m-1}}}\frac{p_{t^n_{j+m}}}{p_{t^n_{j+m-1}}}=1,
\ j=1,\ldots, k_n. \hspace{3ex} \hfill \Box$$


To prove Theorem  \ref{t:asydicalm}, 
the asymptotic distribution of $\Lambda_{k_n}$ 
(see (\ref{eq:distribution})) 
is approximated by the asymptotic distribution
of $W_{k_n}=\sum_{j=1}^{k_n}Y_{n,j}=\sum_{j=1}^{k_n}\tilde Y_{n,j}$ via next lemma that appears in 
Le Cam and Yang (2000); see section 5.2.7, p.92-94, Proposition 8, 
section 5.3 p. 95, conditions
$(B)$ and $(N)$ and p. 108, lines
-10 to -7.


\bel \label{l:supuan}
Assume that under $P_{t_0,n}\footnote{Abuse of notation since $Y_{n,j}=
\tilde Y_{n,j}, \ j=1,\ldots, k_n.$}
$
\begin{equation}
\sup \{|Y_{n,j}|, j=1,\ldots, k_n\} _{n \rightarrow \infty} \longrightarrow 0
\hspace{3ex} \mbox{ in probability. }
\label{eq:supuan}
\end{equation}
When ${\cal A}3$ holds, under $P_{t_0,n}$ 
\[
{\Lambda_{k_n}-2W_{k_n}+\sum_{j=1}^{k_n} Y^2_{n,j}}
_{\hspace{3ex}
n \rightarrow \infty}
\longrightarrow 0
\hspace{3ex} \mbox{ in probability. }
\]
\enl


\bel \label{l:LLC3rd}(see, for example, Le Cam and Yang, 2000, Proposition 1, p. 40 and
p. 41, lines 14-19)
For the binary experiments ${\cal E}_n=\{P_{0,n}, \ P_{1, n}\},$
assume that $\{P_{1,n}\}$ is contiguous to $\{P_{0,n}\}$ and that
the asymptotic distribution of $\log M_n$ under $P_{0,n}$ is
normal with mean $\mu$ and variance $\sigma^2; \ 
M_n=\frac{dP_{1,n}}{dP_{0,n}}$  is the density of the
part of $P_{1,n}$ dominated by $P_{0,n}.$
Then \\
(i) $\mu=-.5\sigma^2,$ and\\
(ii) the asymptotic distribution of $\log M_n$ under $P_{1,n}$ is
normal with mean $.5 \sigma^2$ and variance $\sigma^2.$
\enl

\bel \label{l:contig}
Under ${\cal A}3,$ let $F_{n}$ be the distribution of
$W_{k_n}=\sum_{j=1}^{k_n}Y_{n,j}.$ Assume that $F_{n}$ converges weakly to
a probability distribution $F$ with L\'{e}vy representation
$(\mu, \ \sigma^2,  \ L);$ $\mu$ is the ``shift'', $\sigma^2$ is the
variance of the normal component and
$L$  is the  L\'{e}vy measure. Then,
\\
{\it a)} the sequence $\{P_{T,n}\}$ is contiguous to the sequence $\{P_{t_0,n}\}$
if and only if the variance of $F$ is the limit of the variances
of $F_{n}.$\\
{\it b)} the sequence $\{P_{t_0,n}\}$ is contiguous to the sequence $\{P_{T,n}\}$ if and only if
$L(-1)=0.$
\enl
{\bf Proof.} It follows from the Corollary in Le Cam and Yang (1990, p. 47)
and the Remark, p.48, lines 3-12,
since $P_{t^n_{j-1}}, \ P_{t_j^n}, \ j=1,\ldots, k_n,$
do not have singular parts.
$\hfill \Box$

{\bf Proof of Theorem \ref{t:asydicalm}-Normal distribution for $\log \frac{S_T}{S_{t_0}}$}

Under $P_{t_0,n}$  (defined in (\ref{eq:prodmeas})),
\begin{equation}
E_{P_{t_0,n}}\tilde Y_{n,j}=E_{P_{t^n_{j-1}}}Y_{n,j}
=\int(\sqrt{p_{t_j^n}}\sqrt{p_{t^n_{j-1}}}-1)dP=-h^2(P_{t_j^n},
P_{t^n_{j-1}})=-h^2_{n, j}.
\label{eq:meany}
\end{equation}
The equivalence of $P, \ P_{t^n_{j}}, \ j=0,\ldots, k_n,$ implies
that
\begin{equation}
E_{P_{t_0,n}}\tilde Y^2_{n,j}
=E_{P_{t^n_{j-1}}}Y^2_{n,j}
=\int (\frac{p_{t_j^n}}{p_{t^n_{j-1}}}
-2\sqrt { \frac{{p_{t_j^n}}}{p_{t^n_{j-1}}}}
+1)dP_{t^n_{j-1}}=2h^2_{n, j},
\label{eq:meanysq}
\end{equation}
\begin{equation}
Var_{P_{t_0,n}}(\tilde Y_{n,j})=
Var_{P_{t^n_{j-1}}}(Y_{n,j})=2h^2_{n, j}(1-h^2_{n, j}), \ j=1,\ldots, k_n.
\label{eq:vary}
\end{equation}
For calm stock,  from (\ref{eq:eqecalmexp}) it follows that
$$P_{t_0,n}[\cup_{j=1}^{k_n}\{|\tilde Y_{n,j}|>\epsilon\}]
\le \sum_{j=1}^{k_n}P_{t^n_{j-1}}[|Y_{n,j}|>\epsilon]$$
$$\le {\frac{1}{\epsilon^2}\sum_{j=1}^{k_n}E_{P_{t^n_{j-1}}}Y^2_{n,j}
I(|Y_{n,j}|>\epsilon)}_{n \rightarrow \infty} \longrightarrow 0, \mbox{ and }
$$
\begin{equation}
\sup \{|\tilde Y_{n,j}|, j=1,\ldots, k_n\}
_{n \rightarrow \infty} \rightarrow 0
\label{eq:superuan}
\end{equation}
 in $P_{t_0,n}$-probability. Thus, Lemma \ref{l:supuan} can be used to
approximate the asymptotic distribution of $\Lambda_{k_n}$ (see (\ref{eq:distribution})) with that of 
$W_{k_n}=\sum_{j=1}^{k_n} Y_{n,j}=\sum_{j=1}^{k_n}\tilde Y_{n,j}.$
Assumption ${\cal A}3 (i)$ also holds for calm stocks since
from (\ref{eq:meanysq})
$$2h^2_{n,j}=E_{P_{t^n_{j-1}}}Y^2_{n,j}\le \epsilon^2+E_{P_{t^n_{j-1}}}Y^2_{n,j}
I(|Y_{n,j}|>\epsilon) \le \epsilon^2
+\sum_{j=1}^{k_n}E_{P_{t^n_{j-1}}}Y^2_{n,j}I(|Y_{n,j}|>\epsilon).$$
It follows from ${\cal A}3$ that
$$\sup_{1\le j \le k_n} E_{P_{t_{j-1}^n}}(Y_{n,j}^2)
=\sup_{1\le j \le k_n}h_{n,j}^2{_{\mbox{    }n \rightarrow \infty}}
 \rightarrow 0,$$
and
\begin{equation}
\sum_{j=1}^{k_n} [E_{P_{t_{j-1}^n}}Y_{n,j}]^2
=\sum_{j=1}^{k_n}h_{n,j}^4\le
\sup_{1\le j\le k_n} h_{n,j}^2
\sum_{j=1}^{k_n}h_{n,j}^2{_{\mbox{    }n \rightarrow \infty}} \rightarrow 0.
\label{eq:meansq}
\end{equation}
Since from (\ref{eq:meansq})
and ${\cal A}3 (i)$ the truncated expectation of
$Y_{n,j}$
converges to 0 as $n \rightarrow \infty,$ $j=1,\ldots, k_n,$
(see, for example, Le Cam and Yang, 1990, Lemma 1, p. 34),
and from $A3(ii)$ the sum of the variances is bounded,
for a calm stock the $Y_{n,j}$'s satisfy Lindeberg's
condition and
$W_{k_n}=\sum_{j=1}^{k_{n}} Y_{n,j}$ has distribution $F_n$ that
converges weakly to a normal distribution $F$  with mean $\mu$
and variance $\sigma^2$ (see Le Cam and Yang, 1990, p.44)
\begin{equation}
\mu=-\lim_{n\rightarrow \infty} \sum_{j=1}^{k_n} h_{n,j}^2, \hspace{4ex}
\sigma^2=\lim_{\tau \rightarrow 0} \lim_{n \rightarrow \infty}
\sum_{j=1}^{k_{n}}E_{P_{t^n_{j-1}}}Y^2_{n,j}I(|Y_{n,j}| \le \tau).
\label{eq:variance}
\end{equation}
From Lemma  \ref{l:supuan},
since
$${\sum_{j=1}^{k_{n}}Y^2_{n,j}-
\sum_{j=1}^{k_{n}}E_{P_{t_{j-1}^n}}Y^2_{n,j}}_{n \rightarrow \infty}
\longrightarrow 0$$
in probability,
it follows that
the asymptotic distribution
of $\Lambda_{n}$ is normal
$N(\mu_{[t_0,T]}, \sigma^2_{[t_0,T]}),$ with
\begin{equation}
\mu_{[t_0,T]}=2\mu-\sigma^2, \hspace{4ex} \sigma^2_{[t_0,T]}=4\sigma^2.
\label{eq:meanvarlamda}
\end{equation}
Asymptotic normality
of
$W_{k_n}$ for calm stocks implies
that the limit of the variance of $F_{n}$ is
equal to the variance of $F$ (see Le Cam and Yang, 1990, p. 49, 
lines -20 to -6).
It follows from Lemma  \ref{l:contig}
 {\it a)} that $\{P_{T,n}\}$ is contiguous
to $\{P_{t_0,n}\},$ and from Lemma  \ref{l:LLC3rd} $(i)$
for the mean $\mu_{[t_0,T]}$ and the variance
$\sigma^2_{[t_0,T]}$ of
the asymptotic distribution of $\Lambda_{k_n}$
it holds
$\mu_{[t_0,T]}=-\frac{\sigma^2_{[t_0,T]}}{2}.$
Thus,
from Lemma  \ref{l:contig} {\it b)}, $\{P_{t_0,n}\}$  and $\{P_{T,n}\}$
are contiguous.
Since
$$P_{T,n}[\log \frac{\Pi_{j=1}^{k_n}p_{t^n_j}}
{\Pi_{j=1}^{k_n}p_{t^n_{j-1}}}\le x]=
P_{T,n}[-\log \frac{ \Pi_{j=1}^{k_n}p_{t^n_{j-1}} }
{ \Pi_{j=1}^{k_n} p_{t^n_j} }\le x],$$
(\ref{eq:buyerprob}) follows from $(ii)$ and Proposition  \ref{p:YUequiv}, 
or simply from
Lemma  \ref{l:LLC3rd} $(ii).$


\bel \label{l:probchange}
If a random variable $W$ has under $P$
normal distribution with mean $M$ and variance $\Sigma^2$
and
\begin{equation}
\frac{dP^*}{dP}=e^{AW+C}, \hspace{3ex} -C=MA+\frac{\Sigma^2A^2}{2},
\label{eq:dens1}
\end{equation}
the
distribution of $W$ under $P^*$
is normal with mean $M^*$
and variance $\Sigma^{*2},$
\begin{equation}
M^*=M+A\Sigma^2, \hspace{3ex} \Sigma^{*2}=\Sigma^2.
\label{eq:dens2}
\end{equation}
\enl
{\bf Proof} Follows from the moment generating function $E_{P^*}e^{\rho W}.$ $\hspace{3ex} \hfill \Box$

{\bf Proof of Theorem \ref{t:pri1terms}} Part {\it (i)} follows by taking logarithms in both
sides of the inequality inside the probability, and using Theorem  
\ref{t:infdiv} $(i)$ and
${\cal A}5.$\\
For {\it (ii)} and {\it (iii)}, calculation of expectations is replaced by
calculations of probabilities, after  change of the underlying
probabilities in the expectations.\\
$(ii)$ From ${\bf {\cal A}}5,$ 
mutual absolute
continuity of $P_{t_0,n}$ and $P_{T,n}$ and changing probabilities 
we obtain
$$E_{P_{t_0,n}}[S_{t_0}M_{k_n}
I(M_{k_n}>\frac{X}{S_{t_0}}a^{-1}[t_0,T])|S_{t_0}=s_{t_0}]$$
$$=s_{t_0} 
E_{P_{t_0,n}}
[\Pi_{j=1}^{k_n} \frac{p_{t_j^n}}{p_{t_{j-1}^n}} 
I(\Pi_{j=1}^{k_n} \frac{p_{t_j^n}}{p_{t_{j-1}^n}}>\frac{Xa^{-1}[t_0,T]}
{s_{t_0}})|S_{t_0}=s_{t_0}]$$
$$=s_{t_0}
E_{P_{t_0,n}}\Pi_{j=1}^{k_{n}}
\frac{p_{t_j^{n}}}{ p_{t_{j-1}^{n}} }
I(\Pi_{j=1}^{k_n} \frac{p_{t_j^n}}{p_{t_{j-1}^n}}>
\frac{Xa^{-1}[t_0,T]}{s_{t_0}})
$$
$$=s_{t_0}  
P_{T,n}[
\Pi_{j=1}^{k_n} \frac{ p_{t_j^n} }{ p_{t_{j-1}^n} }
>\frac{Xa^{-1}[t_0,T]}{s_{t_0}}].$$
The result follows from  Theorem \ref{t:infdiv} $(ii)$
by conditioning  the limit
distribution on the value $Y=y$ of the Poissonian component.\\
$(iii)$ From ${\cal A}5,$ 
$$E_{P_{T,n}} [S_{t_0}M_{k_n}
I(M_{k_n}>\frac{X}{S_{t_0}}a^{-1}[t_0,T])|S_{t_0}=s_{t_0}]$$
$$=s_{t_0}
E_{P_{T,n}}
[\Pi_{j=1}^{k_n} \frac{p_{t_j^n}}{p_{t_{j-1}^n}}
I(\Pi_{j=1}^{k_n} \frac{p_{t_j^n}}{p_{t_{j-1}^n}}>\frac{Xa^{-1}[t_0,T]}
{s_{t_0}})|S_{t_0}=s_{t_0}]$$
$$=s_{t_0}
E_{P_{T,n}}\Pi_{j=1}^{k_{n}}
\frac{p_{t_j^{n}}}{ p_{t_{j-1}^{n}} }
I(\Pi_{j=1}^{k_n} \frac{p_{t_j^n}}{p_{t_{j-1}^n}}>
\frac{Xa^{-1}[t_0,T]}{s_{t_0}})
$$
Using Skorohod's theorem, Fatou's Lemma and Theorem \ref{t:infdiv} $(ii),$
$$\liminf E_{P_{T,n}}\Pi_{j=1}^{k_{n}}
\frac{p_{t_j^{n}}}{ p_{t_{j-1}^{n}} }
I(\Pi_{j=1}^{k_n} \frac{p_{t_j^n}}{p_{t_{j-1}^n}}>
\frac{Xa^{-1}[t_0,T]}{s_{t_0}})$$
$$\ge \int \int e^{x+y} I(x>\log \frac{Xa^{-1}[t_0,T]}{s_{t_0}}-y) 
d\Phi(\frac{x+\mu_{[t_0,T]}}{\sigma_{[t_0,T]}})L_T^*(dy).$$
Let $k=\log \frac{Xa^{-1}[t_0,T]}{s_{t_0}}-y.$
For the integral with respect to $x$ 
use (\ref{eq:dens1}) and (\ref{eq:dens2}) with
$M=-\mu_{[t_0,T]}, \ \Sigma^2=\sigma^2_{[t_0,T]},$ to obtain
$e^{-C}=e^{-\mu_{[t_0,T]}+\frac{\sigma^2_{[t_0,T]}}{2}}, \ M^*=-\mu_{[t_0,T]}
+\sigma^2_{[t_0,T]}, \ \Sigma^{*2}=\sigma^2_{[t_0,T]};$
$$\int_k^{\infty} e^x
d\Phi(\frac{ x+\mu_{[t_0,T]} } {\sigma_{[t_0,T]}})=
e^{ -\mu_{[t_0,T]}+\sigma^2_{[t_0,T]} }\int_k^{\infty}
d\Phi(\frac{x+\mu_{[t_0,T]}-\sigma^2_{[t_0,T]}}{\sigma_{[t_0,T]}})$$
$$=e^{-\mu_{[t_0,T]}+\sigma^2_{[t_0,T]} }
\Phi(\frac{-k-\mu_{[t_0,T]}+\sigma^2_{[t_0,T]}}{\sigma_{[t_0,T]}}).$$


$\hspace{3ex} \hfill \Box$

{\bf Proof of Corollary  \ref{c:pri1}} Follows from Theorem  
\ref{t:pri1terms} and 
(\ref{eq:expectedcostatTwithprobdenProduct}) for $\tilde Q_n=P_{t_0,n}, \ P_{T,n}.$
In particular, a translation conditional on the value of the compound Poisson component
in the distribution of $\log \frac{S_T}{S_{t_0}}$ allows to obtain {\it (iii)}.
$\hspace{3ex} \hfill \Box$

{\bf Proof of Theorem  \ref{t:pri1termscalm}} Follows the line of proof 
of Theorem  \ref{t:pri1terms} using
Theorem \ref{t:asydicalm} instead of Theorem  \ref{t:infdiv}.
$\hspace{3ex} \hfill \Box$

{\bf Proof of Corollary  \ref{c:pri1calm} } Follows from Theorem 
\ref{t:pri1termscalm}
and
(\ref{eq:expectedcostatTwithprobdenProduct}) for $\tilde Q_n=P_{t_0,n}, \ P_{T,n}.$
$\hspace{3ex} \hfill \Box$

{\bf Proofs of Theorems and Corollaries in section 5.4}: Follow from those
for section 5.3. $\hspace{3ex} \hfill \Box$

\subsection{Conditions for ${\cal A} 3 $ and 
$\sigma^2_{[t_0,T]}=\sigma^2(T-t_0)$
to hold}

\quad Differentiability conditions are provided below
for ${\cal A} 3 $ to hold. These conditions hold {\it often}
in parametric statistical models.
Let $(\Omega, {\cal F}, P)$ be a probability space
and let $\rho (t), t \in [0,T],$
be a process indexed by $t.$

\bef The process
$\rho$ is differentiable at $\theta$ in $P$-quadratic mean if
there is $U_{\theta},$ its derivative at $\theta,$ such that
\begin{equation}
{\frac{1}{\delta^2}\int[\rho(\theta+\delta)-\rho(\theta)-\delta U_{\theta}]^2
dP}_{\delta \rightarrow 0} \longrightarrow 0.
\label{eq:defqmd}
\end{equation}

When $\theta=0$ (resp. $T$) the limit in (\ref{eq:defqmd})
is taken for $\delta$ positive (resp. negative).
\enf

For the prices-densities $\{p_t, \ t \in [0,T]\},$ let
\begin{equation}
\xi(t)=\sqrt{p_t}, \ t \in [0,T].
\label{eq:xi}
\end{equation}
Then, for the square Hellinger distance $H^2(P_t, P_{\theta})$
of $P_t$ and $P_{\theta}$
 it
 holds
\begin{equation}
H^2(P_t, P_{\theta})=.5\int (\sqrt{p_t}-\sqrt{p_{\theta}})^2dP
=.5\int(\frac{\xi(t)}{\xi(\theta)}-1)^2dP_{\theta}.
\label{eq:defhdxi}
\end{equation}
Conditions for $\xi(t)$ to be quadratic mean
differentiable and examples
of quadratic mean differentiable densities
can be found in Le Cam (1970) and Roussas (1972, Chapter 2).

\bep \label{p:forA3tohold} Assume that
$\xi(t)$ is
$P$-quadratic mean differentiable in
$[t_0,T]$ with derivative $U_{t},$ and that $\sup_{t \in [t_0,T]}
E_P U_{t}^2 <\infty.$
Then, ${\bf {\cal A}}3 $ holds for the densities $p_{t}, \
t \in [0,T].$
\enp


{\bf Proof} For any $\theta$ in $[t_0,T]$ and $\delta$ small,
$$2h^2(P_{\theta+\delta}, P_{\theta})=\int(\frac{\xi(\theta+\delta)}
{\xi(\theta)}-1)^2dP_{\theta}
=\int(\frac{\xi(\theta+\delta)}{\xi(\theta)}-1-\frac{\delta U_{\theta}}
{\xi(\theta)})^2dP_{\theta}\\
$$
$$+ \delta^2
\int (\frac{U_{\theta}}{\xi(\theta)})^2dP_{\theta}+\delta^2 o(1)
=\delta^2 [E_{P_{\theta}}(\frac{U_{\theta}}{\xi(\theta)})^2+o(1)]
=\delta^2[E_PU_{\theta}^2+o(1)].$$
Thus, uniform boundedness  
of $E_PU_{\theta}^2$ implies ${\cal A}3(i)$ holds.

For transaction times with small mesh size in $[t_0, T],$
$$2\sum_{j=1}^{k_n}h^2(P_{t_j^n}, P_{t^n_{j-1}})=\sum_{j=1}^{k_n}(t_j^n-t^n_{j-1})^2
E_PU_{t^n_{j-1}}^2+
o(1)\sum_{j=1}^{k_n}(t_j^n-t^n_{j-1})^2$$
$$\le (T-t_0)^2 [\sup_{t \in [t_0,T]}
E_P U_{t}^2 + o(1)] <\infty.$$
$\hspace{3ex} \hfill \Box$


\bep \label{p:T-t}
Under the assumptions of Proposition  \ref{p:forA3tohold},
$\sigma^2_{[t_0,T]}$ depends on the spacings of the transaction times and
the quadratic mean derivatives of the process $\xi:$\\
a) $$\sigma^2_{[t_0,T]}=4\lim_{\tau \rightarrow 0} \lim_{n \rightarrow \infty}
\sum_{j=1}^{k_n}(t_j^n-t^n_{j-1})^2E_PU_{t^n_{j-1}}^2I(|Y_{n,j}|\le \tau), 
\mbox{ and }$$
b) If $t_j^n-t^n_{j-1}=\delta t^n_{j-1}, \ \delta>0,$ and
$E_PU_{t^n_{j-1}}^2I(|Y_{n,j}|\le \tau)=\frac{c_{\tau}}{t_{j-1}^n}, \ c_{\tau}>0,$
$$\sigma^2_{[t_0,T]}=4\delta (T-t_0)\lim_{\tau \rightarrow 0}c_{\tau}.$$
\enp


{\bf Proof} From the proof of Proposition  \ref{p:forA3tohold},
$$E_{ P_{t^n_{j-1}} }Y_{n,j}^2I(|Y_{n,j}|\le \tau)=
(t^n_j-t^n_{j-1})^2[E_PU_{t^n_{j-1}}^2I(|Y_{n,j}|\le \tau)+o(1)].$$
{\it a)} and {\it b)} follows from (\ref{eq:variancelambda}).
$\hspace{3ex} \hfill \Box$


\begin{thebibliography}{xxx99}




\bibitem{xyz}
Black, F. and Scholes, M. (1973)
The Pricing of Options and Corporate Liabilities.
{\it Journal of Political Economy}, 637-659.

\bibitem{xyz}
Blackwell, D. (1951) Comparison of experiments. {\it Proc. 
2nd Berkeley Symp. Math. Stat. Probab.} 1, 93-102.


\bibitem{xyz}
Brealey, R., Myers, S. and Allen, F. (2008)
{\it Principles of Corporate Finance.} McGraw-Hill, New York.


\bibitem{xyz}
Cox, J. C., Ross, S. A. and Rubinstein, M. (1979)
Option pricing: A simplified approach.
{\it Journal of Financial Economics} {\bf 7}, 229-263.

\bibitem{xyz}
 Detemple, J.  and  Rindisbacher, M.  (2008)
 Dynamic asset liability management with tolerance for limited shortfalls.
{\it  Insurance Math. Econom.} {\bf 43} 281–294.

\bibitem{xyz}
Duffie, D. (2010)  {\it Personal communication}


\bibitem{xyz}
Fama, E. (1970) Efficient Capital Markets: A Review of Theory and Empirical Work. {\it Journal of Finance} 25, 383-417. 


\bibitem{xyz}
Fama, E. (1965) The Behavior of stock market prices. {\it Journal of Business} 38, 34-105. 


\bibitem{xyz}
Geman, H. (1989) The importance of the forward neutral probability
in a stochastic approach of interest rates. Working paper, ESSEC.



\bibitem{xyz}
Jamshidian, F. (1989) An exact bond option formula. {\it Journal of
Finance} {\bf 44}, 205-209.

\bibitem{xyz}
Janssen, A. and Tietje, M. (2013) Applications of the Likelihood Theory in Finance: Modelling and Pricing. {\em Intern. Stat. Rev.}
{\bf 81}, 107-133.





\bibitem{xyz}
Kellison, S. G. (1970) {\it The Theory of Interest.} Irwin Inc., Illinois.

\bibitem{xyz}
Lamberton, D. and Lapeyre, B. (1996)
{\it Introduction to Stochastic Calculus Applied to Finance.}
Chapman and Hall, London.

\bibitem{xyz}
Le Cam, L. M. (1969) {\it Th\'{e}orie Asymptotique de la 
D\'{e}cision Statistique.} Les Presses de l'Universit\'{e} de
Montr\'{e}al.

\bibitem{xyz}
Le Cam, L. M. (1986) {\it Asymptotic Methods in Statistical Decision
Theory.} Springer-Verlag, New York.

\bibitem{xyz}
Le Cam, L. M. (1970) On the assumptions used to prove asymptotic
normality of maximum likelihood estimates. {\it Ann. Math. Stat.} {\bf 41},
802-828.



\bibitem{xyz}
Le Cam, L. M. (1960) Locally asymptotically normal families of distributions. 
{\it Univ. California Publ. Statistics} {\bf 3}, 27-98.

\bibitem{xyz}
Le Cam, L. M. and Yang, G. L. (1990, 2000) 
{\it Asymptotics in Statistics: Some Basic
Concepts.} Springer-Verlag, New York.


\bibitem{xyz}
Lo\`{e}ve, M. (1977) {\it Probability Theory Vol. 1}, 4-th Edition,
Springer-Verlag, New York.

\bibitem{xyz}
Mammen, E. (1986) The statistical information contained in additional
observations. {\it Ann. Stat.} {\bf 14}, 665-678. 

\bibitem{xyz}
Meerschaert, M.  and Scheffler, H.-P.
(2001) {\it Limit Distributions for Sums of Independent
Random Vectors: Heavy Tails in Theory and Practice.} Wiley, New York

\bibitem{xyz}
Merton, R. C. (1976) Option pricing when underlying stock returns are
discontinuous. {\it J. Financial Economics} {\bf 3}, 125-144.

\bibitem{xyz}
Merton, R. C. (1973) Theory of Rational Option Pricing. {\it Bell J. Econ.
and Management Sci.} {\bf 4}, 160-183.

\bibitem{xyz}
Modigliani, F. and Miller, M. H. (1958) The Cost of Capital, Corporate Finance,
and the Theory of Investment. {\it Amer. Econ. Rev.}, {\bf 3}, 261-297.


\bibitem{xyz}
Musiela, M. and Rutkowski, M. (1997) {\it Martingale Methods in 
Financial Modelling.} Springer, Berlin.
\bibitem{xyz}
Pollard, D. (2001) {\it Lectures on Le Cam theory.} 



\bibitem{xyz}
Roussas, G. G. (1972) {\it Contiguous Probability Measures: Some Applications
in Statistics.} Cambridge Univ. Press.


\bibitem{xyz}
Sprenkle, C. (1961) Warrant prices as Indications of Expectations.
{\it Yale Econ. Essays} {\bf 1}, 179-232.


\bibitem{xyz}
Yatracos, Y. G. (2013) 
 Option pricing,  Bayes risk and applications.   http://arxiv.org/abs/1304.5156.
 

\end{thebibliography}
\end{document}